\documentclass[twocolumn]{aastex6}

\usepackage {epsf}
\usepackage{epstopdf}
\usepackage {graphicx}
\usepackage {amssymb,amsmath}

\gdef\Ha{H$\alpha$}
\gdef\Hb{H$\beta$}
\gdef\eqHa{\rm{H}\alpha}
\gdef\eqHb{\rm{H}\beta}
\def\farcs{\hbox{$.\!\!^{\prime\prime}$}}
\def\simgeq{{\raise.0ex\hbox{$\mathchar"013E$}\mkern-14mu\lower1.2ex\hbox{$\mathchar"0218$}}}

\begin {document}

\title {Interpreting the star formation -- extinction relation with MaNGA}

\author{Huan Li\altaffilmark{1,2},
Stijn Wuyts\altaffilmark{3},
Hao Lei\altaffilmark{1},
Lin Lin\altaffilmark{1},
Man I Lam\altaffilmark{1},
M\'ed\'eric Boquien\altaffilmark{4},
Brett H. Andrews\altaffilmark{5},
Donald P. Schneider\altaffilmark{6}
%et al.
} %\ \altaffilmark{*}}

\altaffiltext{1}{Shanghai Astronomical Observatory, Nandan Road 80, Shanghai 200030, China}\email{lihuan@shao.ac.cn (LH); s.wuyts@bath.ac.uk (SW)}\email{haol@shao.ac.cn(HL)}
\altaffiltext{2} {University of Chinese Academy of Sciences, No.19(A) Yuquan Road, Shijingshan District, Beijing, 100049, China}

\altaffiltext{3}{Department of Physics, University of Bath, Claverton Down, Bath, BA2 7AY, UK}

\altaffiltext{4}{Centro de Astronom\'ia, Universidad de Antofagasta, Avenida Angamos 601, Antofagasta 1270300, Chile}

\altaffiltext{5}{PITT PACC, Department of Physics and Astronomy, University of Pittsburgh, Pittsburgh, PA 15260, USA}

\altaffiltext{6}{Department of Astronomy and Astrophysics, The Pennsylvania State University, 525 Davey Laboratory, University Park, PA 16802, USA}

%\altaffiltext{*}{Based on observations collected at the European Organisation for Astronomical Research in the Southern Hemisphere under ESO programmes 092.A-0091, 093.A-0079, 094.A-0217, 095.A-0047, and 096.A-0025.}

\begin{abstract}
We investigate the resolved relation between local extinction and star formation surface density within nearby star-forming galaxies selected from the MaNGA survey. Balmer decrement measurements imply an extinction of the \Ha\ line emission which scales approximately linearly with the logarithm of the star formation surface density: $ A_{\eqHa} = 0.46 \log(\Sigma_{SFR}) + 1.53$. Secondary dependencies are observed such that, at a given $\Sigma_{SFR}$, regions of lower metallicity and/or enhanced \Ha\ equivalent width (EW) suffer less obscuration than regions of higher metallicity and/or lower \Ha\ EW. Spaxels lying above the mean relation also tend to belong to galaxies that are more massive, larger and viewed under higher inclination than average. We present a simple model in which the observed trends can be accounted for by a metallicity-dependent scaling between $\Sigma_{SFR}$ and $\Sigma_{dust}$ via a super-linear Kennicutt-Schmidt relation ($n_{KS} \sim 1.47$) and a dust-to-gas ratio which scales linearly with metallicity ($DGR(Z_{\Sun}) = 0.01$). The relation between the resulting total dust column and observed effective extinction towards nebular regions requires a geometry for the relative distribution of \Ha\ emitting regions and dust that deviates from a uniform foreground screen and also from an entirely homogeneous mixture of dust and emitting sources. The best-fit model features an \Ha\ EW and galactocentric distance dependent fraction of the dust mass in a clumpy foreground screen in front of a homogeneous mixture.
\end{abstract}


\section {Introduction}
\label{intro.sec}

Dust is fundamentally associated to the process of star formation. Dust can serve as a coolant, it can provide shielding from H$_2$-dissociating radiation, and the surface of dust grains can act as a catalyst for the formation of molecular gas, the fuel for star formation. The supernova explosions soon following star formation further contribute to both the formation and destruction of dust \citep[][and references therein]{Popping2017}.

A proper understanding of the impact of dust in terms of its reddening and obscuring effects on the light emitted by gas and stars is also of paramount importance to reliably assess a galaxy's star formation activity. This is particularly true at a spatially resolved level where, for all but the nearest galaxies, maps of the bolometric star formation rate obtained by summing the UV and far-infrared emission remain scarce, and accounting for dust heating by older stellar populations proves non-trivial \citep[see, e.g.,][]{Boquien2016}. While interferometric observations with ALMA or NOEMA increasingly provide such means of bolometrically studying the distribution of star formation, star formation maps for large statistical samples are to date still most efficiently obtained through dust correction methods. The MaNGA integral-field spectroscopic survey \citep{Bundy2015} with its 1-2 kpc scale emission line mapping of several thousands of mass-selected galaxies provides a prime example of this.

From an array of different multi-wavelength diagnostics it is well established that the dust content of galaxies scales with the level of star formation activity. This is the case both when considering measurements of dust mass \citep{daCunha2010} and of dust obscuration \citep{Zoran2006, Wuyts2011, Qinprep}. Popular diagnostics of extinction and star formation that have been used to study such relations at the galaxy-integrated level include IR/UV luminosity ratios and UV+IR bolometric star formation rates (SFR), respectively.  Alternatively, measurements of the Balmer decrement (\Ha/\Hb) and the instantaneous SFR from \Ha\ can be used, for example by exploiting spectroscopy from the SDSS survey which employs large, centrally placed fibers. Having entered the era of large-scale integral-field spectroscopic surveys, such as CALIFA \citep{Sanchez2012}, SAMI \citep{Croom2012} and MaNGA \citep{Bundy2015}, it is an opportune time to extend such studies to subgalactic scales.  

Here, we make use of the exquisite number statistics of MaNGA (internal release MPL-5) to establish the $A_{\eqHa} - \Sigma_{SFR}$ relation at a spatially resolved level, by combining line flux and line ratio measurements of individual spaxels in 977 star-forming galaxies, spanning $\sim$3 dex in stellar mass. In order to interpret the physical origin of this relation, we consider which observables contribute to the observed scatter around the mean relation, and model how dust geometry and metallicity-dependent star formation -- gas -- dust scaling relations contribute to the overall trends followed by the ensemble of 586459 spaxels. As such, our study builds on earlier integral-field spectroscopic work by \citet{Kreckel2013} who investigated the resolved star formation -- extinction relation within 8 nearby galaxies at 100 - 200 pc scales.

After describing our sample selection in Section\ \ref{sample.sec}, we empirically characterize the $A_{\eqHa} - \Sigma_{SFR}$ relation in Section\ \ref{AHa_SIGMAsfr.sec} and investigate its secondary dependencies on other physical parameters in Section\ \ref{dependencies.sec}. We give a brief overview of the ingredients underpinning the observed relation between effective extinction and star formation surface density in Section\ \ref{ingredients.sec}. We illustrate that the observed relation can not be reconciled with a model adopting the simplest dust geometries in Section\ \ref{fixed_models.sec} and next discuss the results of a model with more freedom that is capable of reproducing the observed $A_{\eqHa} - \Sigma_{SFR}$ relation and, at least in a qualitative sense, its secondary dependencies in Section\ \ref{fitted_models.sec}. We summarize our findings and implications for star formation - gas - dust scaling relations and dust geometry in Section\ \ref{summary.sec}.

%\blue{do we need to quote a Chabrier (2003) IMF or cosmology for anything particular?}

\section{Data and sample selection}
\label{sample.sec}

\subsection{The MaNGA survey}

MaNGA is part of the SDSS-IV project \citep{Blanton2017}.  It uses the 2.5 meter telescope at the Apache Point Observatory \citep{Gunn2006} and BOSS spectrographs \citep{Smee2013}.  MaNGA aims to observe $\sim 10,000$ nearby galaxies by 2020 \citep{Bundy2015}, all of them selected from the NASA-Sloan atlas \citep[][\url{http://www.nsatlas.org}]{Blanton2011}.  Fiber bundles of between 19 and 127 fibers are placed on each galaxy, feeding the light to a spectrograph with a wavelength coverage of 360 nm to 1000 nm and spectral resolution of $R \sim 2000$ \citep{Drory2015, Yan2016b, Yan2016a}.  MaNGA's primary sample has a spatial coverage out to 1.5 $R_e$.  The secondary sample, comprising 33\% of the overall sample, covers larger radii up to 2.5 $R_e$ \citep{Wake2017}.

Throughout this paper we use the MaNGA MPL-5 internal release \citep[SDSS-DR14,][]{Abolfathi2017}. \citet{Wake2017} outline the design of the overall MaNGA parent sample and a detailed layout of the reduction pipeline to extract the IFU spectra is presented by \citet{Law2016}. We use advanced products of the MaNGA Data Analysis Pipeline (DAP; \citealt{Westfallinprep}) including emission line and equivalent width measurements carried out on the 3D data cubes after subtraction of the best-fit stellar absorption line spectra using pPXF \citep{Cappellari2004}. Of relevance to our analysis, this means that the DAP measurements of emission line strengths account for underlying stellar absorption. This is particularly important for the \Hb\ emission.

\subsection{Sample selection}

In much of our analysis, we treat the spaxels of all selected galaxies as an ensemble, without reference to which galaxy a given spaxel belongs to. In order to enter the sample, galaxies must be more massive than $\log(M_{star}) > 8.5$ and star-forming. We define the latter criterion by a cut in specific star formation rate at $\log(SSFR) > -11$, corresponding to the minimum of the bimodal star formation rate (SFR) distribution from \citet{Brinchmann2004}.

To guarantee reliable measurements of the Balmer decrement ($\eqHa/\eqHb$), SFR and gas-phase metallicities, derived following the \citet{Pettini2004} O3N2 calibration, we further require selected spaxels to have line detections in all 4 strong optical lines (\Ha, \Hb, [NII], [OIII]) significant at the $> 5\sigma$ level, and line ratios falling within the star-forming branch of the BPT diagram \citep{Kauffmann2003}.  The latter selection also helps to weed out spaxels where photoionization is dominated by hot, low-mass evolved stars, which usually feature LI(N)ER-like ratios in the BPT diagram \citep[see, e.g.,][]{Sarzi2010}.  Recent studies have highlighted considerable contributions from Diffuse Ionized Gas (DIG) to the line emission from nearby galaxies.  While its association with star formation is likely, in certain regimes contributions from evolved stars can play an important role and may affect the accuracy of strong-line metallicity diagnostics \citep{Zhang2017}.  For the O3N2 diagnostic, \citet{Zhang2017} find such biases to the metallicity can be positive or negative in sign, with no systematic shift averaged over the sample they analyse.  \citet{Lacerda2018} formulate EW(\Ha) thresholds below which evolved stars dominate photoionization ($EW(\eqHa) < 3\AA$) and above which star formation dominates entirely ($EW(\eqHa) > 14\AA$).  For intermediate equivalent widths contributions to the photoionization come from both, with potentially a moderate impact on the inferred metallicities.  In our sample, 88\% (99.9\%) of spaxels have $EW(\eqHa) > 14\AA\ (3\AA)$.  This implies that, according to the criteria outlined by \citet{Lacerda2018}, the line emission in the spaxels entering our analysis predominantly traces star-forming complexes.  We tested that eliminating the small fraction of low EW(\Ha) spaxels, contributing to less than 7\% of the overall inferred star formation, does not alter our conclusions.

Finally, in order to avoid the contribution of just a few spaxels from galaxies with faint line emission, we restrict our sample to those galaxies that contain more than 100 spaxels satisfying the above conditions (both the signal-to-noise criterion and the star-forming nature of the ionising source).

%%%%%
% FIG 1
%%%%%
\begin {figure}[t]
%\epsscale{1.0}
\plotone{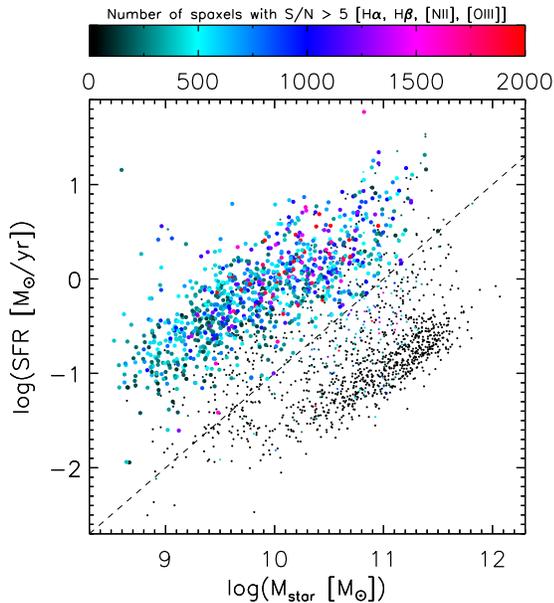}
\caption{Star formation -- stellar mass diagram with all 2778 galaxies from MaNGA MPL-5. Objects are color-coded by the number of spaxels with \Ha, \Hb, [NII] and [OIII] line detections above 5$\sigma$. Larger symbols mark the star-forming galaxies that form part of the sample investigated in this study. The dashed line represents $\log(SSFR) = -11$.}
\label{sample.fig}
%\vspace{0.5cm}
\end{figure}

Overall, this yields a total of 586459 spaxels spread over 977 star-forming galaxies. We present their distribution in the SFR - stellar mass diagram in Figure\ \ref{sample.fig} (large symbols), contrasted to the rest (small symbols) of the underlying MaNGA MPL-5 parent sample from which they were drawn. Here, total star formation rates and stellar masses of the galaxies are taken from the MPA-JHU database\footnote{\url{https://www.mpa-garching.mpg.de/SDSS/DR7/}}, but their values do not enter in the remainder of the paper.  The selected galaxies sample the so-called star-forming main sequence over three orders of magnitude in stellar mass, and span the full distribution of inclinations albeit with a small deficit in highly inclined galaxies compared to what would be expected from random viewing angles (see \citealt{Wake2017} for how the small inclination bias arises from adopting an absolute magnitude rather than stellar mass selection of the MaNGA parent sample).  The MaNGA spaxel size for a typical galaxy in our sample corresponds to $\sim 0.3$ kpc.

\section{The star formation -- extinction relation in MaNGA}
\label{AHa_SIGMAsfr.sec}

%%%%%
% FIG 2
%%%%%
\begin {figure}[t]
%\epsscale{1.0}
\plotone{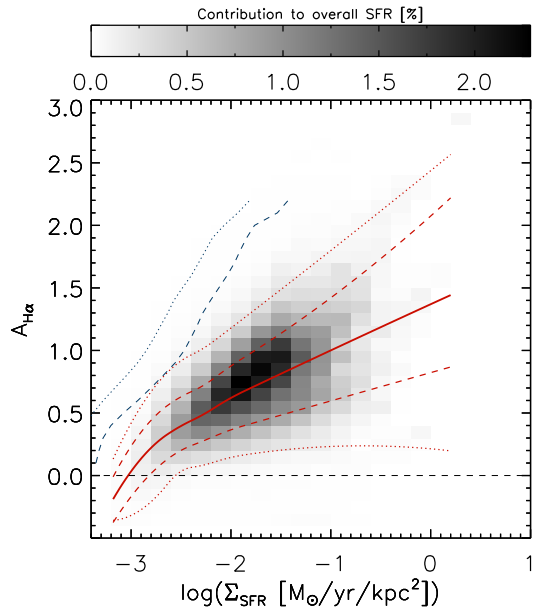}
\caption{The relation between \Ha\ extinction and surface density of star formation for the combined sample of all star-forming spaxels in MaNGA star-forming galaxies. Median extinction levels, as well as central 68th and 95th percentile ranges are marked in red. Blue dashed (dotted) lines indicate 90\% (50\%) completeness limits. Additionally, grayscales illustrate the relative contributions to the summed SFR of all spaxels of different regions in the diagram. A strong relation between extinction and star formation surface density is observed.}
\label{AHa_SIGMAsfr.fig}
%\vspace{0.5cm}
\end{figure}

Figure\ \ref{AHa_SIGMAsfr.fig} presents the distribution of spaxels in the plane of effective \Ha\ extinction $A_{\eqHa}$ versus star formation rate surface density $\Sigma_{SFR}$. Here, $A_{\eqHa}$ is computed following standard methodologies \citep[e.g.,][]{Nelson2016}:
\begin{equation}
A_{\eqHa} = \frac{E(\eqHb - \eqHa)}{k(\lambda_{\eqHb}) - k(\lambda_{\eqHa})} \times k(\lambda_{\eqHa})
\label{AHa.eq}
\end{equation}
where $k(\lambda)$ is adopted from the \citet{Seaton1979} Milky Way reddening curve as fit by \citet{Fitzpatrick1986} and
\begin{equation}
E(\eqHb - \eqHa) = 2.5 \log \left( \frac{(\eqHa/\eqHb)_{obs}}{(\eqHa/\eqHb)_{int}} \right)
\label{EHbHa.eq}
\end{equation}
with the intrinsic ratio $(\eqHa/\eqHb)_{int} = 2.86$ for Case B recombination and $T = 10^4\ K$ \citep{Osterbrock2006}.  The intrinsic ratio depends only weakly on density, with $\eqHa/\eqHb$ ratios confined to values of 2.86 - 2.81 over 4 orders of magnitude in electron density ($n_e = 10^2 - 10^6\ \rm{cm}^{-3}$).  While the dependence on temperature is relatively larger, ranging from 3.04 to 2.75 for 5000 K - 20,000 K \citep{Dopita2003}, the bulk of ionised gas is expected to feature temperatures around 10,000 K such that the anticipated scatter in observed $\eqHa/\eqHb$ ratios due to variations in the intrinsic Balmer decrement are trumped by dust effects and do not impact our analysis.

The same extinction measure is applied to dust correct the \Ha-based SFR of a given spaxel following \citet{Kennicutt1998}, which for a \citep{Chabrier2003} IMF scales as:
\begin{equation}
SFR_{\eqHa} [M_{\odot}/\rm{yr}] = 4.68 \times 10^{-42}\ L_{\eqHa}\ e^{A_{\eqHa}/1.086},
\end{equation}
with the surface density of star formation in the spaxel following naturally from the spaxel's angular extent ($0\farcs 5 \times 0\farcs 5$) and known galaxy distance. We note that no inclination correction is applied to the surface densities and the $\Sigma_{SFR}$ values in Figure\ \ref{AHa_SIGMAsfr.fig} thus correspond to projected surface densities. When modeling the relation in Section\ \ref{ingredients.sec} and beyond, the deprojected surface density enters in some steps (notably when applying the Kennicutt-Schmidt relation), but not in others. E.g., for extinction the total (projected) column of dust matters.

Figure\ \ref{AHa_SIGMAsfr.fig} visualizes the relation between extinction and local star formation activity in two ways:  in red we mark the median $A_{\eqHa}$, as well as the central 68th and 95th percentile intervals.  These are derived as a function of $\log(\Sigma_{SFR})$ using the COnstrained B-Splines nonparametric regression quantiles method (COBS; \citealt{Ng2015}), which combines spline regression with likelihood-based knot selection and quantile regression.  The ensemble of spaxels is in number heavily dominated by low surface brightness and low $\Sigma_{SFR}$ regions constituting the outskirts of galaxies, in the range $-3 \lesssim \log(\Sigma_{SFR}) \lesssim -2$. While these spaxels account for a significant fraction of the areal coverage above the signal-to-noise thresholds outlined in Section\ \ref{sample.sec} they only represent a minor contribution to the total star formation rate of the galaxies in our sample. In order to give an indication of which spaxels contribute most significantly to the total star formation budget we therefore weigh the gray-scale coding of ($A_{\eqHa}$, $\Sigma_{SFR}$) bins in Figure\ \ref{AHa_SIGMAsfr.fig} by the fractional contribution of the bin to the summed star formation rate of all spaxels. It is evident that this fractional contribution is dominated by spaxels in the range $-2 \lesssim \log(\Sigma_{SFR}) \lesssim -1$. The fact that in detail the grayscaled distribution is slightly offset upwards compared to the binned medians can be understood from the MaNGA sample definition \citep{Wake2017}. Larger, more massive galaxies, which tend to feature more obscuration, are targeted at somewhat higher redshifts. Due to the different physical pixel scales the same $\Sigma_{SFR}$ then translates to a larger amount of star formation within the spaxels of these more distant systems.

Since the \Hb\ line emission is fainter than \Ha, and more so when extinction levels are high (see Eq\ \ref{AHa.eq} - \ref{EHbHa.eq}), incompleteness due to \Hb\ being fainter than 5$\sigma$ is anticipated to affect the high $A_{\eqHa}$, low $\Sigma_{SFR}$ region of the diagram first. To assess its impact, we compute the 90\% (blue dashed line) and 50\% (blue dotted line) completeness curves as follows. We reduce each spaxel in our sample in brightness while keeping the relative line ratios fixed, until one of the four lines required for our analysis (\Ha, \Hb, [NII], [OIII]) no longer satisfies the $S/N > 5$ criterion. This corresponds to shifting each spaxel to lower $\Sigma_{SFR}$ by a factor which depends on the significance of its line detections. Next, we assess in slices of $A_{\eqHa}$ down to which $\Sigma_{SFR}$ we would retain 90\% (50\%) of the spaxels. The resulting curves are indicated in Figure\ \ref{AHa_SIGMAsfr.fig} with blue dashed and dotted lines, respectively. We conclude that the distribution of $A_{\eqHa}$ is not abruptly cut off by incompleteness effects, and therefore proceed by analyzing the $A_{\eqHa} - \Sigma_{SFR}$ relation as observed.

The observed trend can be approximated by a linear relation between $A_{\eqHa}$ and $\log(\Sigma_{SFR})$:
\begin{equation}
A_{\eqHa} = (0.46 \pm 0.03) \log(\Sigma_{SFR}) + (1.53 \pm 0.01),
\label{linear_fit.eq}
\end{equation}
with the slope and intercept determined through linear regression to the individual data points. Errors on the coefficients are derived from 1000 bootstrap iterations in which mock samples are drawn with replacement from the original sample of galaxies whose spaxels enter the analysis. The normalized median absolute deviation (NMAD) of individual spaxels from the linear fit given in Equation\ \ref{linear_fit.eq} is limited to NMAD = 0.24 mag, with a gradual increase in the width of the central 68th percentile in $A_{\eqHa}$ by a factor of $\sim2.5$ over 3 orders of magnitude in $\Sigma_{SFR}$.

It is worth emphasizing that in constructing Figure\ \ref{AHa_SIGMAsfr.fig}, the Balmer decrement enters on both axes, as it was used to derive the dust-corrected star formation surface density as well as $A_{\eqHa}$. In the absence of any dust correction (i.e., when simply scaling the observed $\Sigma_{\eqHa}$ to $\Sigma_{SFR,\ uncorrected}$ by applying a constant factor $4.68 \times 10^{-42}$) spaxels span a reduced dynamic range in star formation surface density and the relation between $A_{\eqHa}$ and $\log(\Sigma_{\eqHa})$ is no longer linear. Instead, the median $A_{\eqHa}$ shows a relatively shallow increase up to $\log(\Sigma_{SFR,\ uncorrected}) \sim -1.5$ beyond which it saturates. The scatter around the median relation is enhanced, as also quantified by reduced Spearman rank and Pearson correlation coefficients of $R \approx 0.41$, compared to $R \approx 0.62$ for the $A_{\eqHa} - \Sigma_{SFR}$ relation captured by Equation\ \ref{linear_fit.eq} where the total (i.e., dust-corrected) star formation surface density is considered.

A second consideration is that, in converting the Balmer decrement to the \Ha\ extinction, a reddening law was invoked (Equation\ \ref{AHa.eq}). As stated previously, a Milky Way reddening law is adopted throughout this paper. Here, we note that adoption of the \citet{Calzetti2000} reddening law established based on observations of starburst galaxies (or variations thereof accounting for extra extinction to nebular regions) would yield a steeper relation as $k(\eqHa) / [k(\lambda_{\eqHb}) - k(\lambda_{\eqHa})] = 2.61$ compared to 1.90 for the Milky Way law. For completeness, the best-fit relation adopting a \citet{Calzetti2000} law is $A_{\eqHa} = 0.65 \log(\Sigma_{SFR}) + 2.10$. We checked that the conclusions from our modeling of the star formation - extinction relation are robust to the choice of reddening law as long as one adopts one law consistently throughout the analysis. 

%%%%%
% FIG 3
%%%%%
\begin{figure*}
\centering

{
 \includegraphics[width = .32\linewidth]{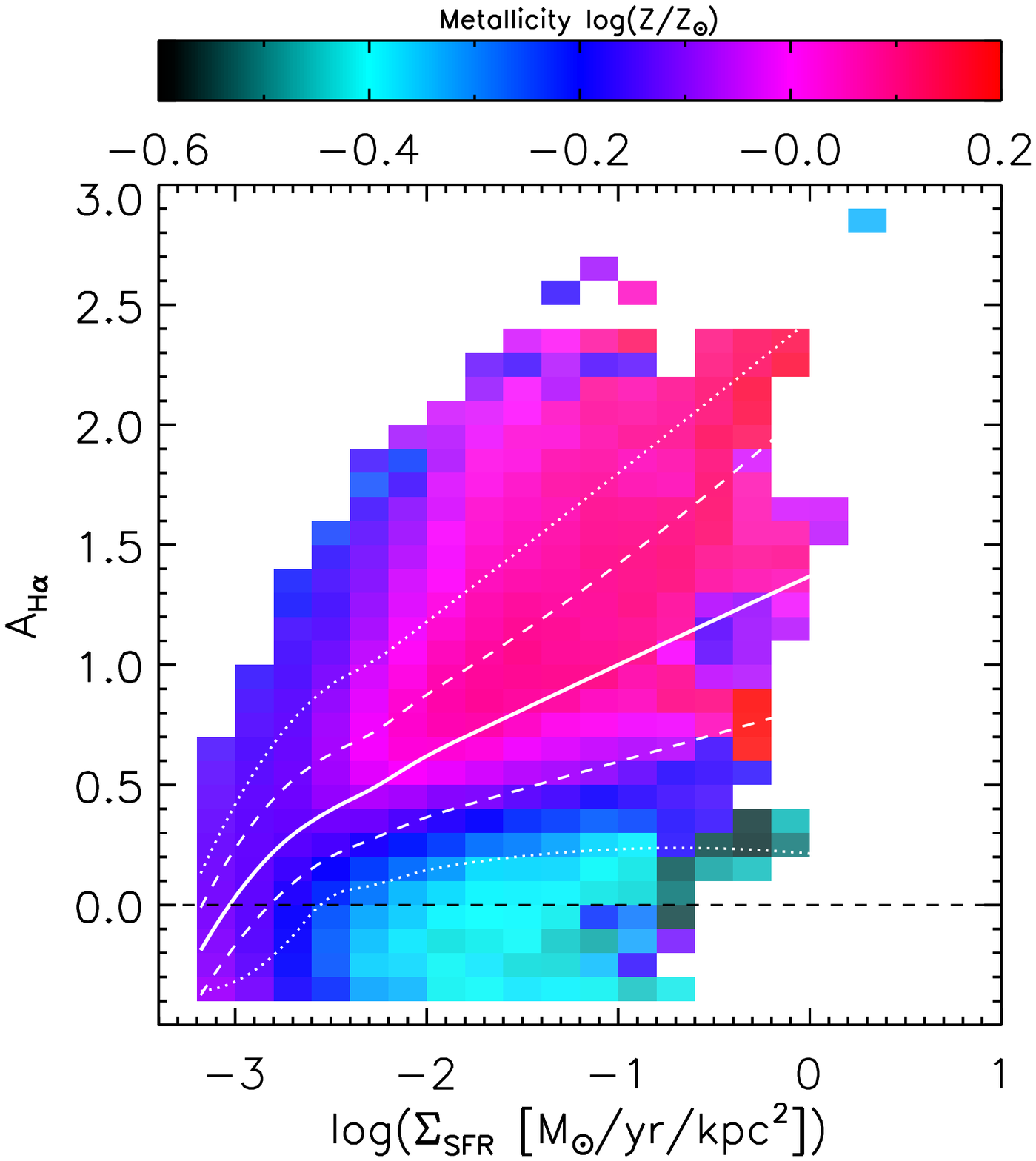}
 \includegraphics[width = .32\linewidth]{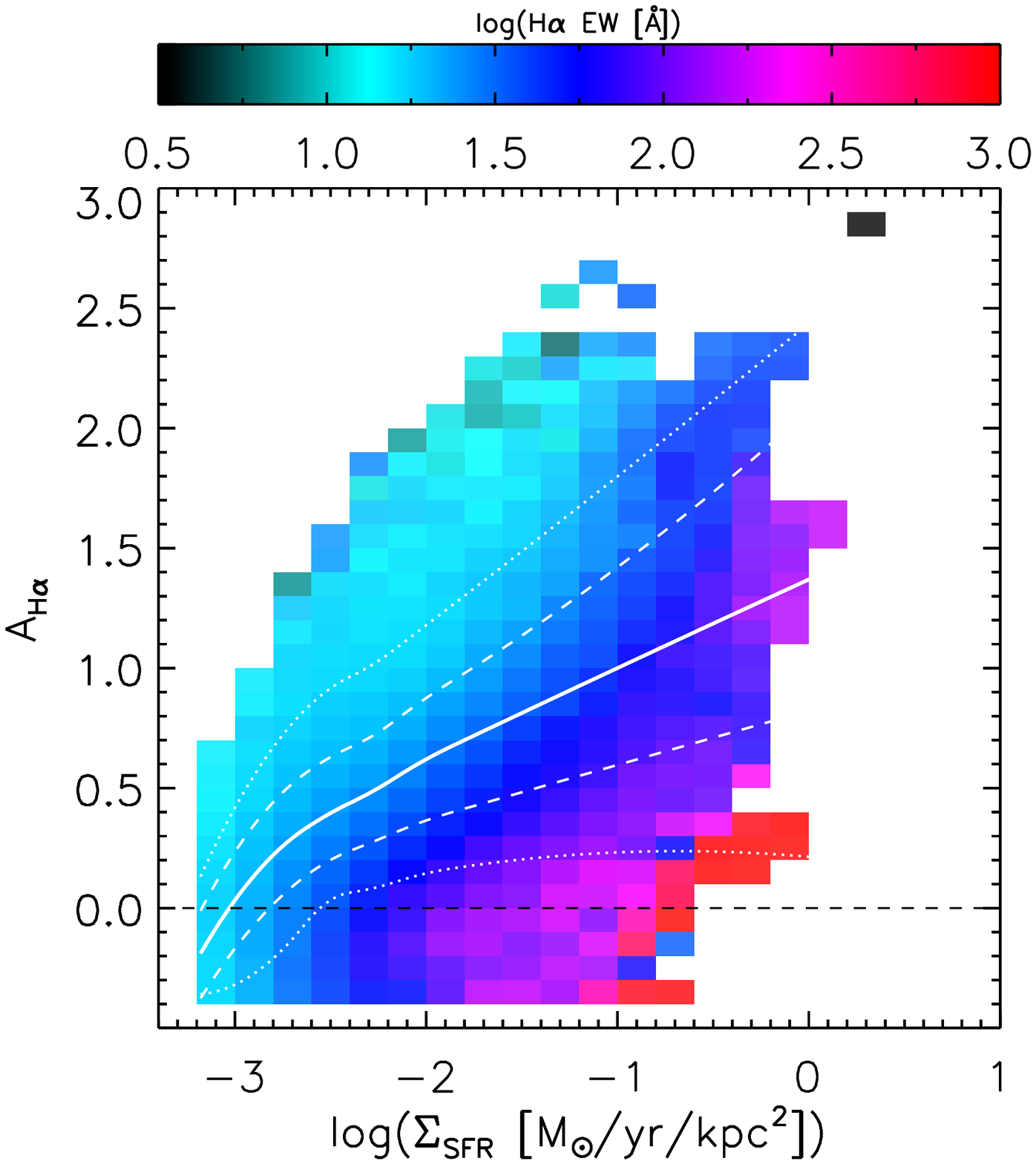}
 \includegraphics[width = .32\linewidth]{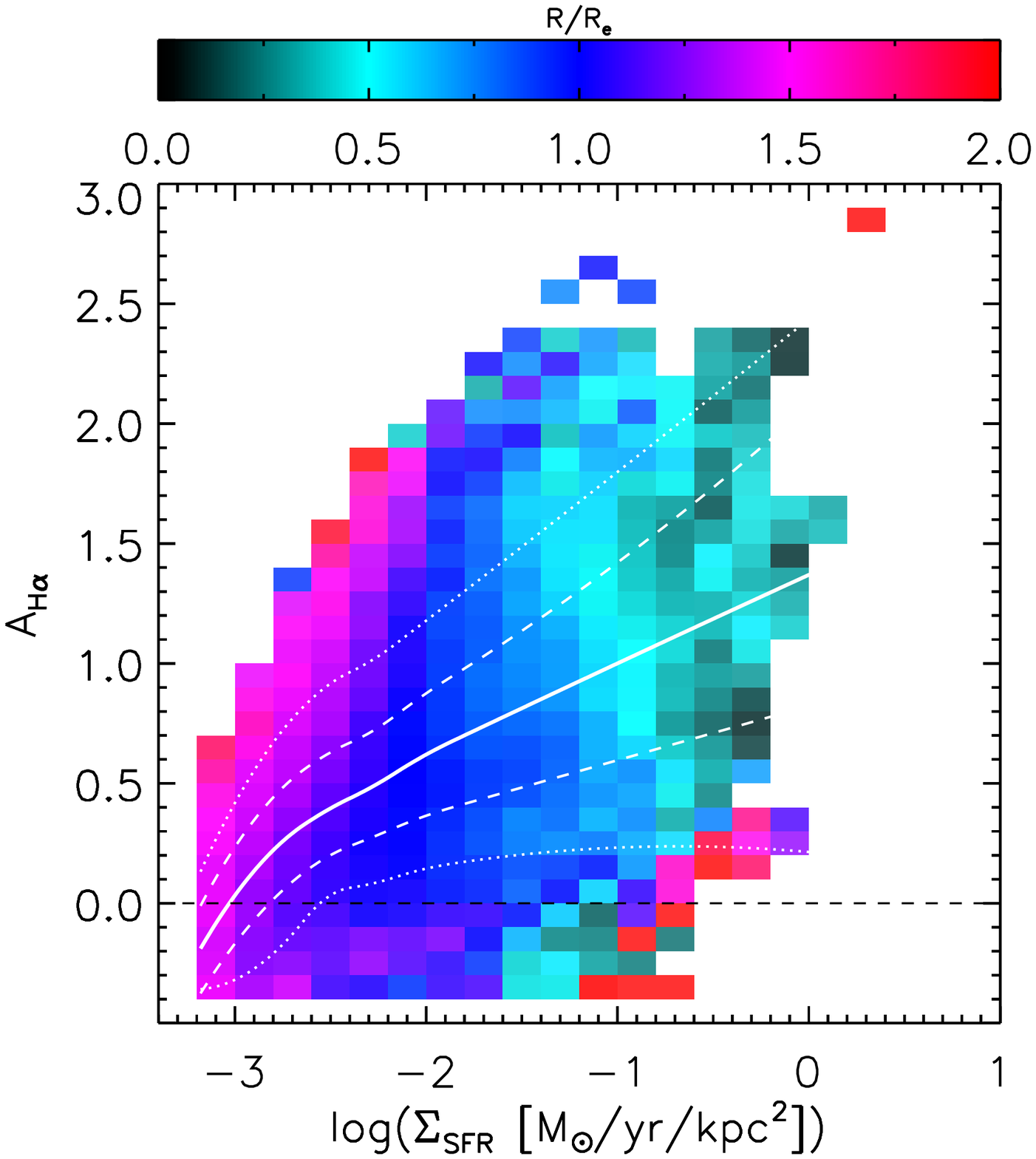}
 }

 {
 \vspace{2mm}
 \includegraphics[width = .32\linewidth]{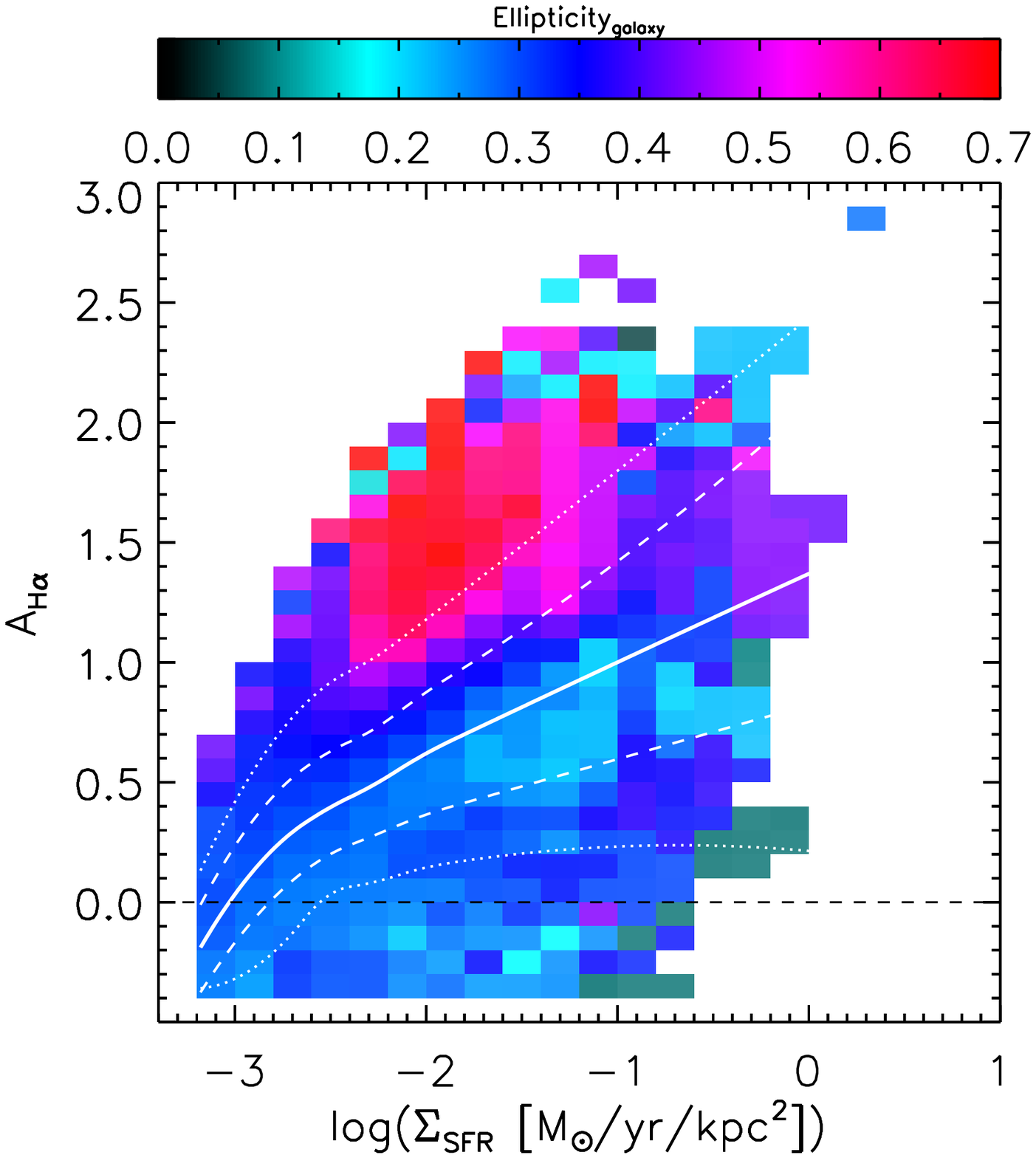}
 \includegraphics[width = .32\linewidth]{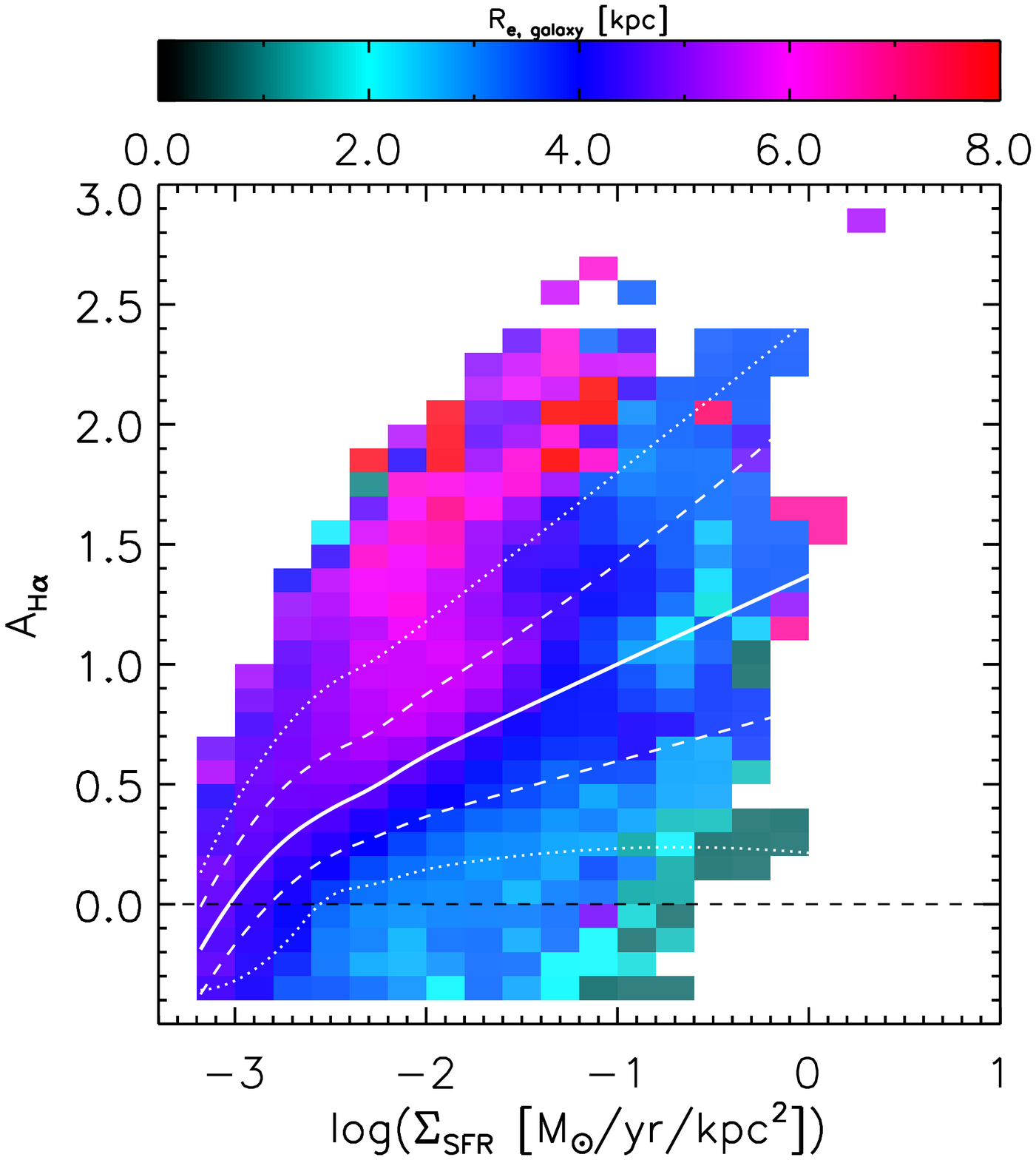}
 \includegraphics[width = .32\linewidth]{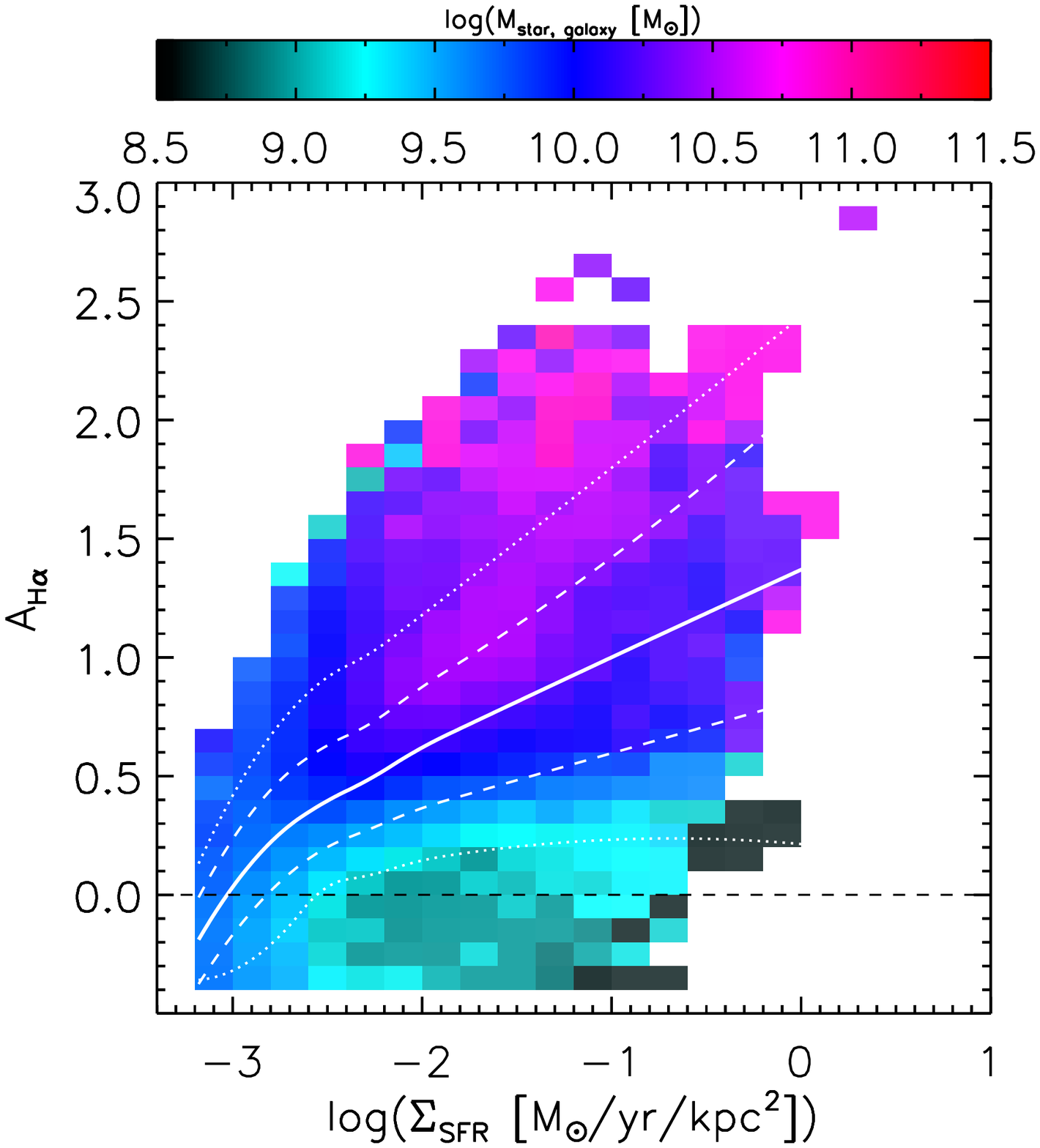}
}
 
\caption{The observed star formation -- extinction relation color-coded by the median gas-phase metallicity ({\it top left}), \Ha\ equivalent width ({\it top middle}) and galactocentric distance ({\it top right}) of spaxels in the bin, and ellipticity ({\it bottom left}), effective radius ({\it bottom middle}) and stellar mass ({\it bottom right}) of the galaxies to which the spaxels belong.}
\label{dependencies.fig}
\end{figure*}

Finally, we stress that, while providing a reasonable empirical description of the median $A_{\eqHa}$ as a function of star formation surface density, the functional form of Equation\ \ref{linear_fit.eq} is not physically motivated. We will explore more physically motivated models in the same diagram in Sections\ \ref{fixed_models.sec} - \ref{fitted_models.sec}, but first consider whether residuals from the median relation relate to any other observables.

\section{Secondary dependencies}
\label{dependencies.sec}

\subsection{Second-parameter dependencies for entire sample}
Whereas the bulk of spaxels populates a relatively tight $A_{\eqHa} - \Sigma_{SFR}$ relation (R = 0.62; NMAD = 0.24) the full distribution exhibits a significant range in extinction levels at any given $\Sigma_{SFR}$, particularly in the high surface density tail of the distribution ($\log(\Sigma_{SFR}) > -1$), where the 95th percentile range spans 1.7 - 2 mag. This prompts the question whether the scatter around the relation is random or alternatively finds its origin in dependencies on secondary parameters other than $\Sigma_{SFR}$. Here, we will distinguish between potential secondary parameters at the spaxel level and properties of the galaxies to which the spaxels belong.    

A first quantity of interest that can be determined on a per spaxel basis is the gas-phase metallicity.  We employ the O3N2 calibration by \citet{Pettini2004}, as the same recipe has also been frequently used in the gas and dust scaling relations we will turn to in Section\ \ref{ingredients.sec}. The physical relevance stems from the fact that the amount of star formation (the x-axis of Figures\ \ref{AHa_SIGMAsfr.fig} - \ref{dependencies.fig}) is known to depend on the available amount of fuel (i.e. the cold gas, or more specifically molecular gas reservoir), whereas the dust responsible for extinction (the y-axis of Figures\ \ref{AHa_SIGMAsfr.fig} - \ref{dependencies.fig}) is composed of heavy elements. Indeed, the anticipated trend is seen: spaxels lying below the median relation in Figure\ \ref{dependencies.fig} ({\it top left panel}) contain gas that is relatively devoid of metals. Conversely, the outliers towards high extinction values are more typically characterized by high gas-phase metallicities.  The observed trend is in line with results obtained by \citep{Boquien2013} who studied the UV and IR properties of 4 resolved and 27 unresolved galaxies with gas and metallicity measurements, and report a similar metallicity dependence of the relation between optical depth and gas surface density.

In a similar manner we explore and find secondary dependencies with \Ha\ equivalent width ({\it top right panel of Figure\ \ref{dependencies.fig}}).  Contrasting the strength of gaseous line versus stellar continuum emission, the equivalent width can be thought of as a proxy for specific star formation rate, albeit with potential influences due to differential dust extinction depending on dust geometry \citep[see, e.g.,][]{Calzetti2000, Wild2011, Wuyts2013, Price2014}. Empirically, we find that spaxels with more than average extinction correspond to regions of lower \Ha\ EW than counterparts of the same $\Sigma_{SFR}$ that are less affected by dust.  Turned around, at a given $\Sigma_{SFR}$ spaxels of enhanced \Ha\ EW (i.e., with a higher ratio of ongoing star formation relative to assembled stellar mass) effectively suffer from relatively less extinction.

Along the locus encompassing the bulk of the spaxels (marked by the 95th percentiles) the median galactocentric distance of spaxels in a given bin, expressed in normalized units $R/R_e$), changes mostly with $\Sigma_{SFR}$, not with $A_{\eqHa}$.  This reflects the galaxies' SFR profiles, which are on average radially declining.

Turning to the physical properties of the galaxies the spaxels belong to, we investigate secondary trends with ellipticity, galaxy size and stellar mass in the remaining panels of Figure\ \ref{dependencies.fig}. It is immediately apparent that the more extreme outliers above (below) the median $A_{\eqHa} - \Sigma_{SFR}$ relation come from galaxies with high (low) ellipticities.  Considering star-forming galaxies as axi-symmetric disks, ellipticity is a measure of inclination.  The observed trends can thus readily be interpreted as more inclined systems featuring a larger projected column of obscuring material and hence more extinction. We note that inclining a galaxy further not only moves its spaxels up in the diagram, but also to the right, as we are plotting the observed (i.e., projected) surface densities. If we were to show deprojected star formation surface densities on the x-axis instead, the slope of the best-fit linear relation would become somewhat shallower, with little change in zero point or scatter: $A_{\eqHa} = 0.38 \log(\Sigma_{SFR,\ deprojected}) + 1.43$ (NMAD = 0.26 mag).

Finally, the spaxels of large ($\gtrsim 5$ kpc), massive ($\gtrsim 10^{10.5}\ M_{\odot}$) galaxies tend to lie predominantly above the median relation. Of course, these trends may be strongly coupled to the aforementioned ones. E.g., the existence of a stellar mass - metallicity relation implies more massive galaxies have a more enriched Inter-Stellar Medium (ISM) and hence plausibly more dust and extinction. In fact, as we will demonstrate in Section\ \ref{fitted_models.sec}, a simple model can be constructed that reproduces, at least in a qualitative sense, the secondary dependence on galaxy stellar mass without using the stellar mass itself as input information.

\subsection{Consideration of subsamples}
As a side note, we remind the reader that the galaxy sample from which our ensemble of spaxels is drawn comprises both galaxies from the MaNGA primary sample, accounting for two thirds of the objects and reaching out to 1.5 $R_e$, and from the secondary sample covering to larger radii.  Likewise, while the bulk of the sample (65\% of galaxies) features spatial resolutions in the range 1 -- 2 kpc, 32\% of objects are mapped at a resolution better than 1 kpc and 11\% at a resolution shallower than 2 kpc (see \citealt{Wake2017} for details of the MaNGA sample selection).  

We repeated the analysis of the star formation -- extinction relation and its secondary dependencies for each of the subsamples (primary/secondary; three bins of spatial resolution).  The same secondary dependencies are observed for all subsamples, with slight nuances.  E.g., by construction the MaNGA secondary sample contributes more spaxels located at large $R/R_e$.  In common between all subsamples is a near-linear trend between $A_{\eqHa}$ and $\log(\Sigma_{SFR})$ which tends to zero extinction around surface densities of $\log(\Sigma_{SFR}) \sim -3$.  Slight differences in slope of the relation are observed for the different subsamples, from 0.50 to 0.42 for the primary versus secondary sample, and from 0.39 for the subset of galaxies with $< 1$ kpc resolution to a slope of 0.50 for the subset observed at $> 2$ kpc resolution.  We note, however, that this is not an assessment of how the star formation -- extinction relation would change if resolving the same galaxies to a higher or lesser degree.  Instead, the modest slope changes for the different subsamples largely stem from the second parameter dependencies discussed in this section coupled with the fact that, per MaNGA sample definition and for reasons of optimal use of fiber bundles, the more massive galaxies are typically observed at a shallower spatial resolution in physical (kpc) units \citep{Wake2017}.

\section{Ingredients to a simple model}
\label{ingredients.sec}

Having discussed the empirical results obtained from the MaNGA data cubes, we here briefly introduce the physical ingredients we will use to link the resolved star formation and extinction using a simple model.

\subsection{Kennicutt-Schmidt relation}
\label{KS.sec}

The Kennicutt-Schmidt (KS) relation is characterized by a zero point and slope describing the dependency of the star formation per unit area on the surface density of gas \citep{Kennicutt98}. It is now well established that this relation, which has been the focus of ample studies at the galaxy-integrated and spatially resolved level, is strongest when considering only the gas component in the molecular phase (as opposed to the total cold gas reservoir including the atomic phase; see, e.g., \citealt{Bigiel08}). This conclusion even holds in the HI-dominated outer disk regions (\citealt{Bigiel2010}; \citealt{Schruba2011}; and see, e.g., \citealt{Fu2010} for model implementations.). 

Here, we follow \citet{Bigiel08} in defining the zero point $ZP_{KS}$ as the $\log(\Sigma_{SFR})$ for regions with a molecular gas surface density of $10\ M_{\Sun}\ \rm{pc}^{-2}$, where the data are most constraining:
\begin{equation}
\log(\Sigma_{SFR}) = n_{KS} \log \left(\frac{\Sigma_{molgas}}{10\ M_{\Sun}\ pc^{-2}} \right) + ZP_{KS}.
\label{KS.eq}
\end{equation}
Using the HERACLES nearby galaxy survey these authors find $ZP_{KS} = -2.23 \pm 0.20$ (including the standard $\sim 36\%$ correction for helium) and a linear slope $n_{KS} = 1.0 \pm 0.2$.  However, it is worthwhile to point out that the precise slope of the KS relation remains a topic of debate in the literature, with results varying from the sub-linear ($n_{KS} \sim 0.8$; e.g., \citealt{Shetty2013}), near-linear ($n_{KS} \sim 1 - 1.2$; e.g., \citealt{Bigiel08, Genzel2010, Tacconi2013}), to super-linear regime ($n_{KS} \sim 1.3 - 1.8$; e.g., \citealt{Momose2013}). 

\subsection{Dust-to-gas ratio}
\label{DGR.sec}

It is well established that the dust-to-gas ratio (DGR) of galaxies, or subregions within galaxies, depends on metallicity. In fact, the dust-to-metal ratio shows little variation, implying a linear relation of the form:
\begin{equation}
\log(DGR) = \log(DGR(Z_{\Sun})) + \log(Z/Z_{\Sun})
\label{DGR.eq}
\end{equation}
where the Solar oxygen abundance is $(12 + \log(O/H))_{\Sun} = 8.67$ \citep{Asplund04}.  Measurements of the dust-to-gas ratio at Solar metallicity $DGR(Z_{\Sun})$ range from 0.0062 \citep{Remy2014} to 0.01 \citep{Leroy2011, Sandstrom2013}, with at least some of this range stemming from systematic uncertainties in metallicity calibrations.  The reference work by \citet{Draine2011} quotes $DGR(Z_{\Sun}) = 0.0091$.

It deserves some consideration that what in Equation\ \ref{DGR.eq} is referred to as gas relates to the total amount of cold gas, including both the molecular and atomic phase. That is at least the case for sight lines to regions with significant star formation. Comparisons between dust- and CO-based cold gas masses imply that some but not all of the galaxy-integrated HI mass is traced by dust \citep[see, e.g.,][]{Bertemes2018}, but the outer HI disks devoid of star formation, enriched material and dust are by construction not part of the spaxels entering our analysis.

In computing the dust surface density in our toy model, we will assume a maximum HI contribution corresponding to the empirical metallicity-dependent surface density threshold for conversion to the molecular phase observed by \citet{Leroy2008}, \citet{Bigiel08} and \citet{Wong2013}:
\begin{equation}
\log(\Sigma_{HI}) = \log(10\ M_{\Sun}\ \rm{pc}^{-2}) + \log(Z/Z_{\Sun}).
\label{HI.eq}
\end{equation}
The above relation accounts for contributions from helium, and is consistent with model predictions by \citet{McKee2010} and \citet{Sternberg2014}. Motivated by the observational findings on dust, molecular and atomic gas from \citet{Bertemes2018} we further impose the HI surface density not to exceed the molecular gas surface density. This approach to adding in HI contributions when converting from gas to dust is favored by the data presented in \citet{Bertemes2018}, and also yields better matching model predictions in our present analysis than obtained when following the more extreme bracketing approaches of adding in no HI at all, or always the maximum amount dictated by Equation\ \ref{HI.eq}.

\subsection{Dust geometry}
\label{geometry.sec}

Combining the star formation - gas (Section\ \ref{KS.sec}) and gas - dust (Section\ \ref{DGR.sec}) scaling relations, we obtain an estimate of the dust surface mass density. Following \citet{Kreckel2013} this dust column can be translated to an extinction under the assumption of a uniform foreground screen through the linear relation:
\begin{equation}
A_{V, screen}(\Sigma_{dust}) = 0.67 \frac{\Sigma_{dust}}{10^5\ M_{\Sun}\ \rm{kpc}^{-2}}\ \rm{mag}
\label{foreground_screen.eq}
\end{equation}
or, converted to the extinction at the wavelength of \Ha:
\begin{equation}
A_{\eqHa, screen} = A_{V, screen} \frac{k(\lambda_{\eqHa})}{R_{V, MWG}}
\end{equation}
where $R_{V, MWG} = 3.1$ corresponds to the Milky Way value of $R_V \equiv \frac{A_V}{E(B-V)}$.

Note that in Equation\ \ref{foreground_screen.eq}, the extinction scales linearly with surface density, as opposed to linearly with its logarithm as we saw in Equation\ \ref{linear_fit.eq}.  A potentially non-linear mapping from $\Sigma_{SFR}$ to $\Sigma_{dust}$ by means of a super-linear KS relation is by itself insufficient to explain this change of functional form.

Instead, we have to recognize that the configuration of a uniform foreground screen may be physically implausible, and that other dust geometries have to be invoked. To this end, \citet{Calzetti94} introduced three qualitatively different scenarios \citep[see also][]{Natta84}: (1) a uniform foreground screen, (2) a clumpy foreground screen, and (3) a homogeneous mixture of dust and emitting sources. Scenario (1) can be considered a limiting case of scenario (2) in which the average number of clumps along the line of sight $N_{clumps}$ is large.  By default, we assume scenarios (1) and (2) to correspond to a screen sufficiently far from the emitting sources such that the effects from scattering in the foreground screen are negligible.  Where relevant, we comment on how alternative assumptions, from isotropic to an-isotropic or forward-only scattering within the screen would affect our results (see Equations 12 - 17 in \citep{Calzetti94}).

In our default model the dust geometry will be captured by two quantities: the aforementioned $N_{clumps}$ specifying the clumpiness of the foreground screen and a parameter $f_{screen}$ which represents the fraction of the dust in a foreground screen component. The remaining dust, i.e. a fraction $(1 - f_{screen})$ of the total column, is taken to be homogeneously mixed with the emitting sources (in our case the HII regions from which the \Ha\ emission originates). For the homogeneous mixture component, we follow \citet{Calzetti94} in adopting their prescription for a wavelength-dependent anisotropy of the scattering (their equations 15, 16, 21b and 22).

We now briefly review the essential characteristics of dust geometries (2) and (3), referring the reader to \citet{Calzetti94} for more details.

For a foreground screen composed of a Poissonian distribution of clumps, the average total optical depth $\tau(\lambda) = 0.921 A(\lambda)$ can be written as the sum of the optical depths of all clumps along the line of sight (assumed to be of equal properties):
\begin{equation}
\tau(\lambda) = N_{clumps} \tau_{clump}(\lambda).
\label{tau.eq}
\end{equation}
Due to the clumpiness of the dust distribution, the effective optical depth $\tau_{eff} \equiv - \ln(I/I_0)$ which describes the ratio of emerging ($I$) and originally emitted radiation ($I_0$) will be lower than that:
\begin{equation}
\tau_{eff, clumpy\ screen}(\lambda) = N_{clumps} \left[ 1 - e^{-\tau_{clump}(\lambda)} \right].
\label{tau_eff.eq}
\end{equation}
A Taylor expansion of Equation\ \ref{tau_eff.eq} shows that in the limit of large $N_{clumps}$ the effective optical depth approaches the expression for total optical depth in Equation\ \ref{tau.eq}, effectively describing the uniform foreground screen we started with. More generally, we can write the effective extinction towards \Ha\ for any value of $N_{clumps}$ as
\begin{eqnarray}
A_{\eqHa, clumpy\ screen} & = & 1.086 \left[ e^{-\tau_{clump}(\eqHb)} - e^{-\tau_{clump}(\eqHa)}\right] \notag \\
 & & \times N_{clumps} \times \frac{k(\eqHa)}{k(\eqHb) - k(\eqHa)}
\end{eqnarray}
with $\tau_{clump}$ relating to the total dust column $\tau$ as prescribed by Equation\ \ref{tau.eq}.

For a homogeneous mixture, the effective optical depth is also reduced compared to the uniform foreground screen, such that the emitted radiation is attenuated by a factor
\begin{equation}
\gamma(\lambda) \equiv \frac{1 - e^{-\tau(\lambda)}}{\tau(\lambda)}.
\end{equation}
Translated to an effective extinction towards \Ha, this yields:
\begin{eqnarray}
A_{\eqHa, mix} & = & 1.086\ \ln{\left[ \frac{\gamma(\eqHa)}{\gamma(\eqHb)} \right]} \notag \\  
 & & \times \frac{k(\eqHa)}{k(\eqHb) - k(\eqHa)}
\end{eqnarray}

% tauB = tau_b - tau_a = ln( (Ha/Hb) / 2.86 )
% 
% A_Ha = 2.5 * log( HaHb / 2.86 ) * k(Ha) / (k(Hb) - k(Ha))

% A_Ha = A_V k(Ha) / R_V

%A_Ha = 2.5 * log( e^tauB ) * k(Ha) / (k(Hb) - k(Ha))
%          = 2.5 * log(e) * tauB * k(Ha) / (k(Hb) - k(Ha))
%          = 1.086 * tauB * k(Ha) / (k(Hb) - k(Ha))
 
The above finally yields an overall effective extinction, resulting from the combination of a component of dust in a homogeneous mixture and a component in a clumpy foreground screen, of the following form:
\begin{equation}
A_{\eqHa, eff} = A_{\eqHa, mix} + A_{\eqHa, clumpy\ screen}
\end{equation}
where $A_{\eqHa, mix}$ is computed starting from a dust column of $(1 - f_{screen}) \Sigma_{dust}$ and $A_{\eqHa, clumpy\ screen}$ is computed starting from a dust column of $f_{screen} \Sigma_{dust}$.

%%%%%
% FIG 4
%%%%%
\begin {figure}[t]
%\epsscale{0.33}
\epsscale{1.0}
\plotone{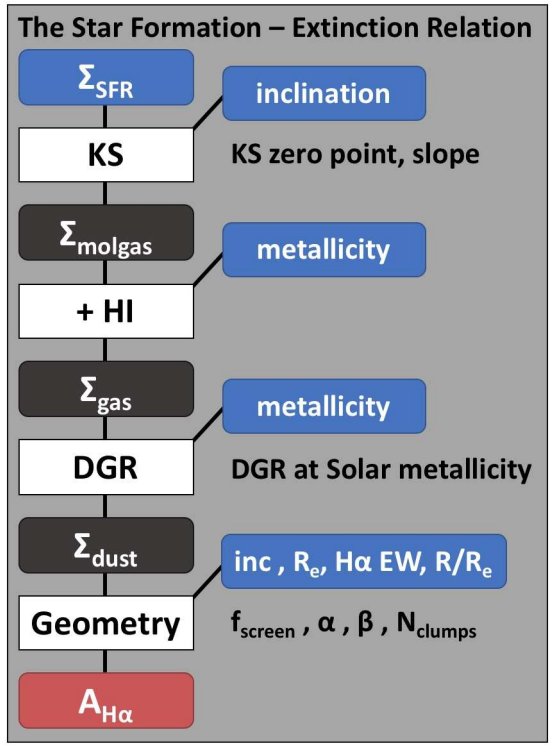}
%\epsscale{1.0}
\caption{Overview of our simple model relating the resolved star formation surface density to the local extinction inferred from the Balmer decrement.  Input observables are marked in blue. Steps involving the application of gas (Kennicutt-Schmidt; KS) and dust (Dust-to-Gas Ratio; DGR) scaling relations as well as an analytical dust geometry describing a combination of a homogeneous mixture and (clumpy) foreground screen are indicated with white rectangles. The associated parameters which are left free or on which priors are imposed appear to the right of the respective conversion step. The model is tuned by contrasting the predicted to the observed $A_{H\alpha}$ on a spaxel-by-spaxel basis.}
\label{cartoon.fig}
%\vspace{0.5cm}
\end{figure}

\subsection{Summary of modeling approach}
\label{summary_ingredients.sec}

In the remainder of this paper, we will explore a set of models based on the physical ingredients outlined in this Section, aimed at reproducing the observed $A_{\eqHa} - \Sigma_{SFR}$ relation and its secondary dependencies. We build this up initially exploring simpler models that fail, to various degrees, to capture the richness of information in the observed data set, eventually reaching our favored model which is visualized in Figure\ \ref{cartoon.fig}. Here, observed properties used as input are indicated in blue. These are the observed $\Sigma_{SFR}$ of all spaxels, their metallicity, \Ha\ EW and normalized galactocentric radius, and the inclination and size of the galaxies they belong to. These inputs are combined through a set of scaling relations and a dust geometry (white rectangles in Figure\ \ref{cartoon.fig}) to obtain a predicted $A_{\eqHa}$ for every of the 586459 spaxels.

The galaxy inclination, for example, enters in two conversion steps of our model.  First, to deproject the observed $\Sigma_{SFR}$ in order to apply the KS relation, which is defined in the plane of the disk.  Next, in the geometry step to translate the dust surface density in the plane of the disk via a projected dust column to an effective extinction.  The number of clumps along the line of sight of a given spaxel also depends on the galaxy inclination, expressed in the form of the observed axial ratio $b/a$ as:
\begin{equation}
N_{clumps} = N_{clumps,\ norm}\ \left(\frac{R_e}{5\ \rm{kpc}}\right)\ \left(\frac{b}{a}\right)^{-1}.
\end{equation}
Here, the normalization factor $N_{clumps,\ norm}$ is the number of clumps along the line of sight for a galaxy of 5 kpc (the typical size of galaxies in our sample) when seen face on ($\frac{b}{a} = 1$). Larger galaxies, or galaxies viewed under a more inclined angle, will have a larger number of clumps along the line of sight, making the foreground screen approximate more closely a uniform foreground screen. As detailed in Section\ \ref{geometry.sec}, the latter is more effective than a clumpy foreground screen at attenuating the emitted light for a given total dust column. It is the normalization factor $N_{clumps,\ norm}$ that is the parameter we will fit for, and for brevity we have dubbed this parameter $N_{clumps}$ in Figure\ \ref{cartoon.fig} and all subsequent Sections.

While we start out exploring models where the same fixed $f_{screen}$ value applies to all spaxels, assuming such a universal ratio of dust in foreground and mixture components may not be realistic.  Various studies in the literature document variations in the relative geometry of dust and gas as a function of star formation activity \citep[e.g.,][]{daCunha2010} and/or location in the galaxy \citep[e.g.,][]{Liu2013, Tomicic2017}.  As we will describe in Section\ \ref{fitted_models.sec}, also our own analysis prompts us to introduce additional freedom by expressing in our favored model the dust -- gas geometry as dependent on the \Ha\ EW and spaxel position within the galaxy such that
\begin{equation}
f_{screen} = f_{screen,\ norm}\ 10^{\alpha [\log(\eqHa\ EW) - 1.4]}\ 10^{\beta [R/R_e]}
\label{fscreen.eq}
\end{equation}
with $\alpha$, $\beta$ and $f_{screen,\ norm}$ being the free parameters. Versions of the model lacking the extra EW and $R/R_e$ dependent factors in Equation\ \ref{fscreen.eq} (i.e., adopting a universal $f_{screen}$ for all spaxels) already capture the bulk of observed trends but require an implausibly steep slope of the KS relation and feature systematic residuals when contrasting the model predictions to observations. The parameter $f_{screen,\ norm}$, representing the fraction of dust in a foreground screen for a central spaxel of typical EW ($\log(\eqHa\ EW) = 1.4$), will for brevity be referred to as $f_{screen}$ when presenting our favored model in Section\ \ref{fitted_models.sec}, keeping the parameter names consistent also with the simpler model versions without $\alpha$ and $\beta$ discussed in Section\ \ref{fixed_models.sec}.

Returning to our model overview in Figure\ \ref{cartoon.fig}, dark gray boxes denote the intermediate physical quantities ($\Sigma_{molgas}$, $\Sigma_{gas}$, $\Sigma_{dust}$) and parameters involved in the various steps are listed next to the respective white rectangles. In our favored model (Section\ \ref{fitted_models.sec}) we fit the model to the data using the Levenberg -- Marquardt algorithm to minimize
\begin{equation}
\chi^2 = \sum_{i=1}^{N_{spaxels}} \frac{\left( A_{\eqHa\ observed,\ i} - A_{\eqHa\ model,\ i}\right)^2}{\sigma_{A_{\eqHa\ observed,\ i}}^2}
\end{equation}
with the 5 parameters $n_{KS}$, $f_{screen}$, $\alpha$, $\beta$ and $N_{clumps}$ left entirely free, and the KS zero point and DGR at Solar metallicity merely allowed to vary within the range of measurements in the literature: $-2.33 < ZP_{KS} < -2.13$ and $-2.3 < \log(DGR) < -2.0$. We then evaluate our best-fit model visually by considering the equivalents of Figure\ \ref{AHa_SIGMAsfr.fig} and\ \ref{dependencies.fig}, and additionally by checking for any systematic dependencies of the residuals $A_{\eqHa\ model} - A_{\eqHa\ observed}$ on the various observables.

\section{Naive (and failing) models}
\label{fixed_models.sec}

%%%%%
% FIG 5
%%%%%
\begin {figure}[t]
%\epsscale{0.49}
\plotone{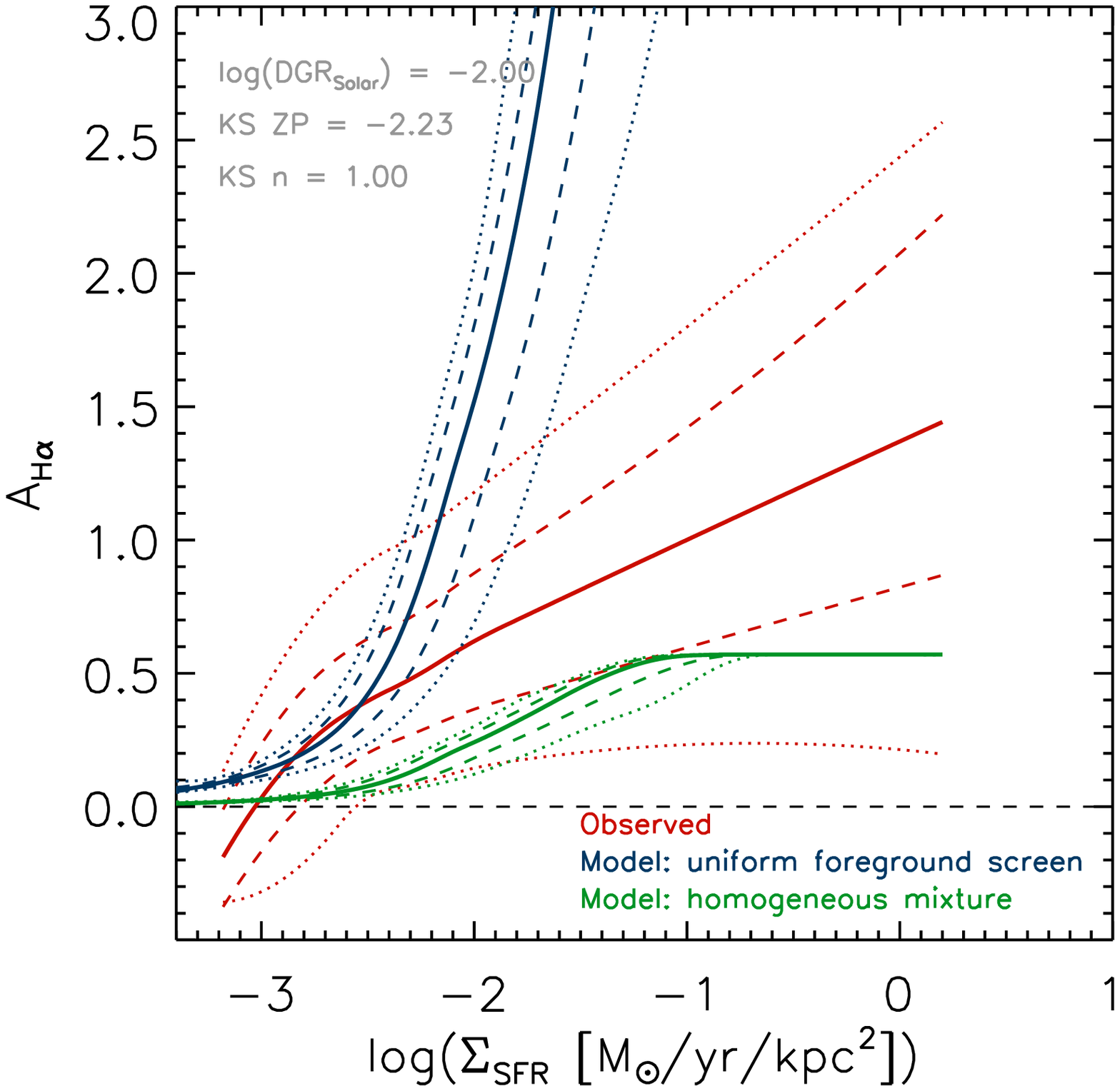}
\plotone{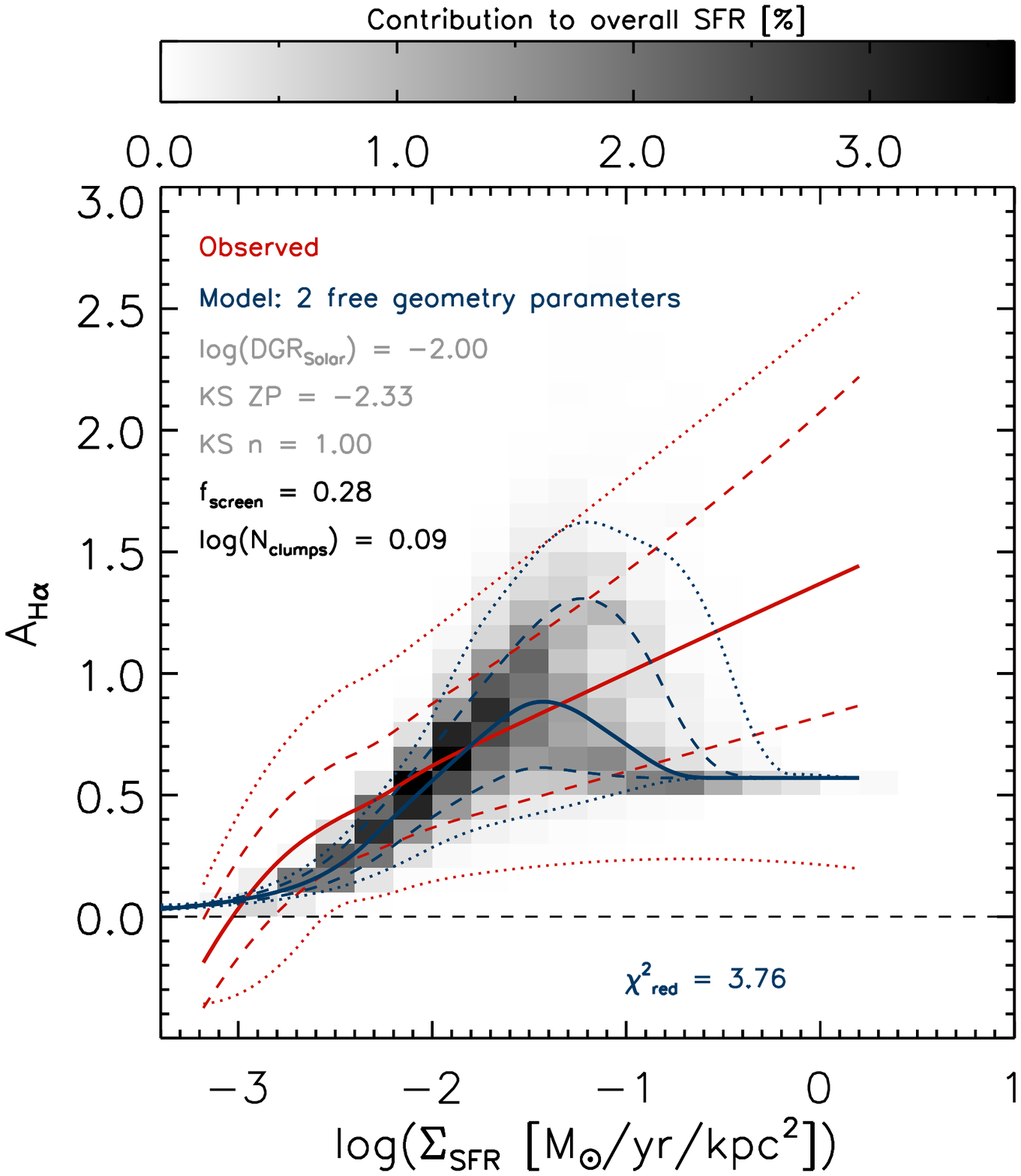}
\caption{{\it Top:} The observed star formation - extinction relation ({\it quantile statistics in red taken from Figure\ \ref{AHa_SIGMAsfr.fig}}) contrasted to predictions from simple models assuming the dust is distributed as a uniform foreground screen ({\it blue curves}) or homogeneously among the \Ha\ emitting regions ({\it green curves}).  Parameters describing the star formation and gas-to-dust scaling relations are kept fixed. Neither dust geometry provides a satisfying description of the observed relation.
{\it Bottom:} The observed star formation - extinction relation ({\it red curves}) contrasted with the best-fit two-parameter model allowing a range of dust geometries ({\it blue curves}).  Grayscales correspond to the spaxel distribution as predicted by the model. Listed are best-fit values of $f_{screen}$ and $\log(N_{clumps})$, whereas parameters related to the dust -- gas -- star formation scaling relations are constrained to canonical values from the literature (see text).}
%\blue{I.e. use loadct,4,/silent and color=50, 100, 200 for the various naive model results.}}

\label{naive_models.fig}
%\vspace{0.5cm}
\end{figure}

In a first attempt to interpret the observed star formation - extinction relation of Figure\ \ref{AHa_SIGMAsfr.fig}, we compute for each spaxel the anticipated $A_{\eqHa}$ starting from its (dust-corrected) $\Sigma_{SFR}$ and metallicity under the assumption of dust geometry (1), the uniform foreground screen. That is, we fix $f_{screen} = 1$, assign a large value of $N_{clumps} = 1000$ and for simplicity adopt canonical values for $\log(DGR(Z_{\Sun})) = -2$, $ZP_{KS} = -2.23$ and $n_{KS} = 1$, although as discussed in Section\ \ref{ingredients.sec} not all of them are free of debate.  As such, there are no free parameters to fit.

Contrasting the blue and red curves in the top panel of Figure \ref{naive_models.fig}, it is immediately apparent that the approximation of a uniform foreground screen provides a poor description of the observed star formation - extinction relation.  The model prediction does not reproduce the observed roughly linear relation between $A_{\eqHa}$ and $\log(\Sigma_{SFR})$ but instead exhibits extinction levels that rise increasingly rapidly as one moves to higher $\log(\Sigma_{SFR})$. This overshooting effect is inherent to the adopted geometry (see Equation\ \ref{foreground_screen.eq} where the linear surface density enters, not its logarithm). We verified that models with all dust in a uniform foreground screen remain of poor quality even when allowing freedom in the parameters characterizing the star formation -- gas ($ZP_{KS}$, $n_{KS}$) and gas -- dust ($DGR(Z_{\Sun})$) relations.

Following similar arguments, we can rule out that all dust is homogeneously mixed with the line-emitting sources (i.e., dust geometry (3) parameterized as $f_{screen} = 0$).  This is illustrated by the green curve in the top panel of Figure\ \ref{naive_models.fig}.  The net effect of a homogeneous mixture is that the effective extinction saturates, leading to a paucity of spaxels with $A_{\eqHa} > 0.6$, even at large star formation and hence gas and dust columns, in clear contrast to the observations.

Having established that the simplest bracketing scenarios for dust geometry are inadequate, consistent with findings by \citet{Kreckel2013}, we now introduce more freedom in our model introducing the two geometry parameters $f_{screen}$ and $N_{clumps}$ from Section\ \ref{ingredients.sec}. In other words, we explore the effect of allowing a dust distribution that is composed in part of a homogeneous mixture and in part a clumpy foreground screen. Since we are now fitting, we also allow $ZP_{KS}$ and $DGR(Z_{\odot})$ to vary slightly, but only within the small interval spanned by independent observational results quoted in the literature ($-2.33 < ZP_{KS} < -2.13$ and $-2.3 < \log(DGR) < -2.0$). The slope of the star formation law $n_{KS}$ is kept fixed to unity, for exploration at a later stage (Section\ \ref{fitted_models.sec}). The resulting model is presented in the bottom panel of Figure\ \ref{naive_models.fig}. The best-fit $f_{screen}$ corresponds to a case intermediate between a foreground screen and a homogeneous mixture, and the best-fit screen is clumpy rather than uniform. Despite the additional freedom and reduced chi-squared value of the fit that improved sensitively to $\chi^2_{red} = 3.76$ it can be observed that the improvement stems mostly from getting the typical extinction right in the range $\log(\Sigma_{SFR}) \sim -1.5$ where spaxels contribute the bulk of the overall star formation. Considering the full dynamic range in $\Sigma_{SFR}$ it is clear that the model does not reproduce the shape of the overall star formation -- extinction relation quantified in Equation\ \ref{linear_fit.eq}. The model therefore remains of poor quality.

\section{Fitting the $A_{\eqHa} - \Sigma_{SFR}$ relation}
\label{fitted_models.sec}

%%%%%%%%%%% Figure 6 %%%%%%%%%%%%%%%%%%%
\begin{figure}[b]
%\epsscale{0.8}
\includegraphics[width = .98\linewidth]{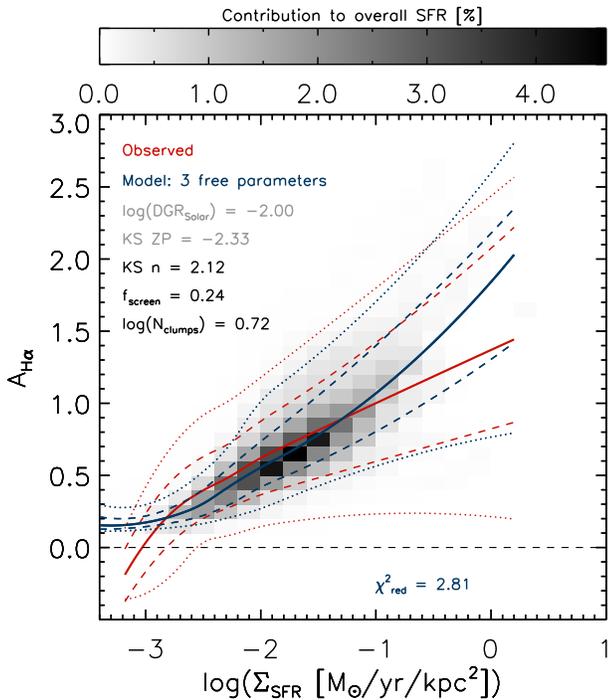}
\caption
{The observed star formation - extinction relation ({\it red curves}) contrasted with the best-fit three-parameter model ({\it blue curves and grayscales}). Listed are best-fit values of $f_{screen}$, $log(N_{clumps})$ and KS slope ($n_{KS}$).  Other parameters related to the dust - gas - star formation scaling relations are left to vary only within a narrow interval of literature results based on independent observations.}
\label{without_alphabeta.fig}
\end{figure}
%%%%%%%%%%%%%%%%%%%%%%%%%%%%%%%%%%%%%%%%%%%%

%%%%%%%%%%%%% Fig 7 %%%%%%%%%%%%%%%%%%%%%%
%\begin{figure}[t]
%\epsscale{0.8}
%\plottwo{eps/plot_onemodel_without_alphabeta1_Z.eps}{eps/plot_onemodel_without_alphabeta1_HaEW.eps}
%\plottwo{eps/plot_onemodel_without_alphabeta1_RRe.eps}{eps/plot_onemodel_without_alphabeta1_Ellipticity.eps}
%\plottwo{eps/plot_onemodel_without_alphabeta1_Re.eps}{eps/plot_onemodel_without_alphabeta1_Mstar.eps}

{
\begin{figure*}%[t]
\centering
{
 \includegraphics[width = .32\linewidth]{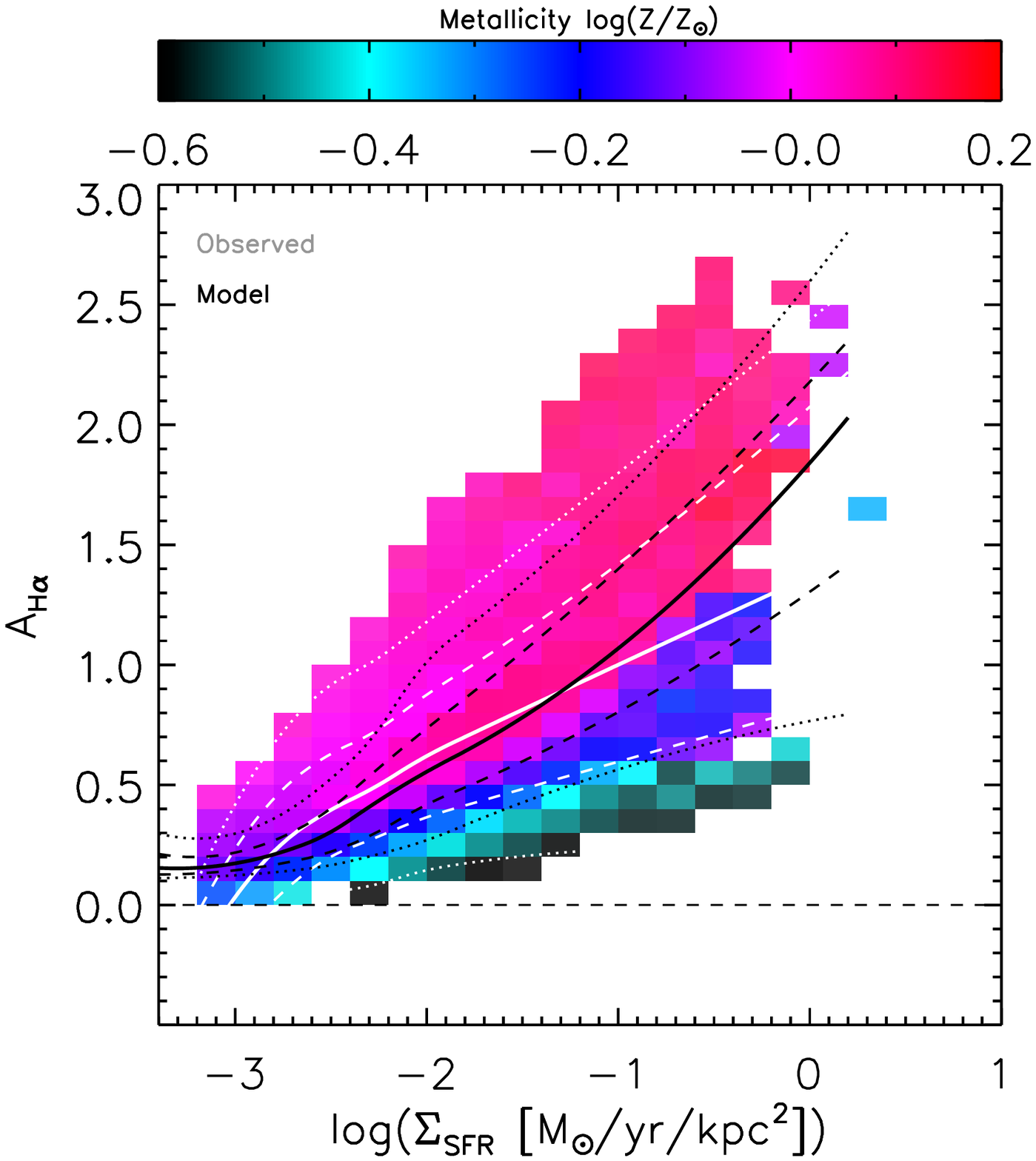}
 \includegraphics[width = .32\linewidth]{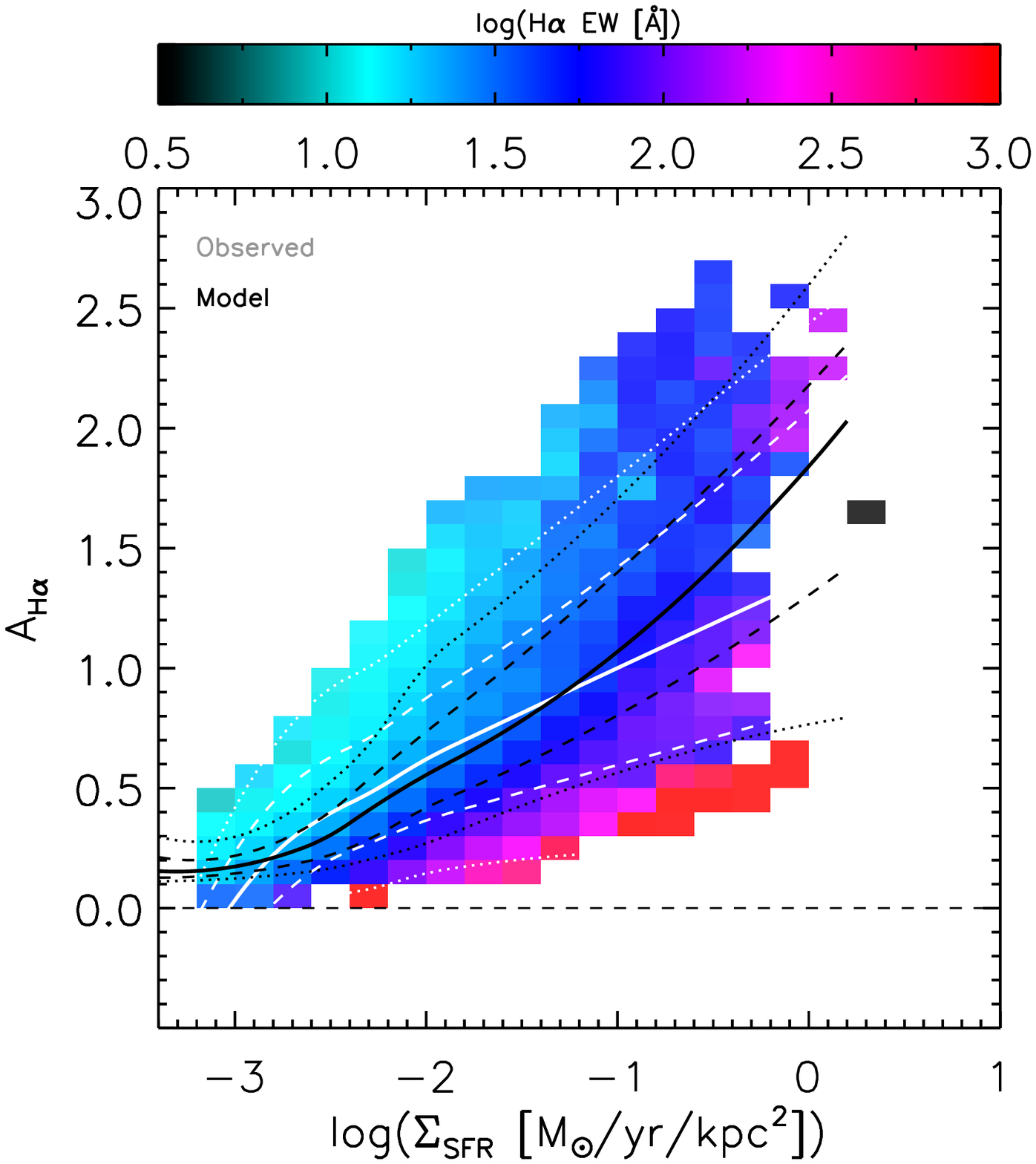}
  \includegraphics[width = .32\linewidth]{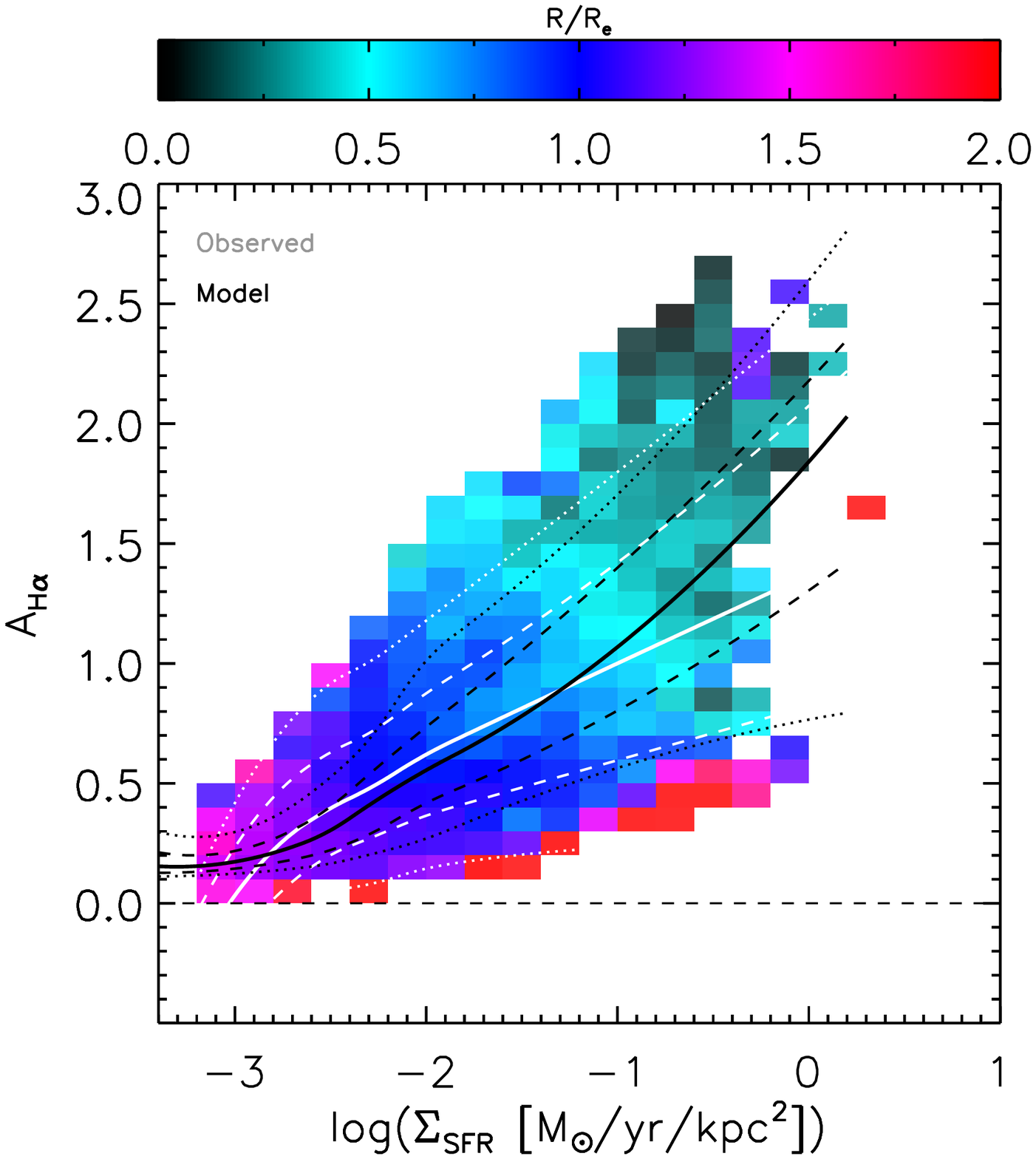}
}

{
\vspace{2mm}
 \includegraphics[width = .32\linewidth]{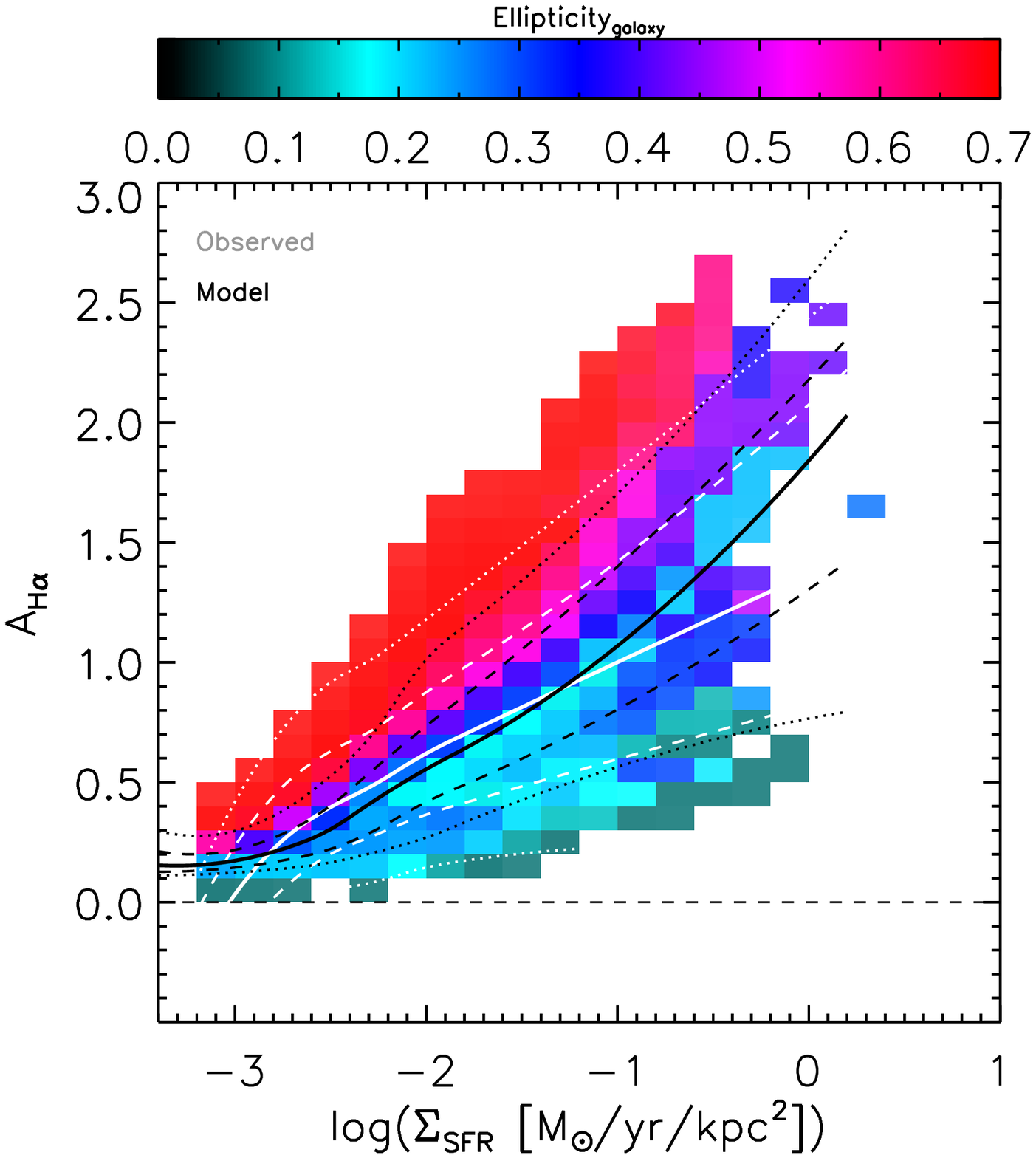}
 \includegraphics[width = .32\linewidth]{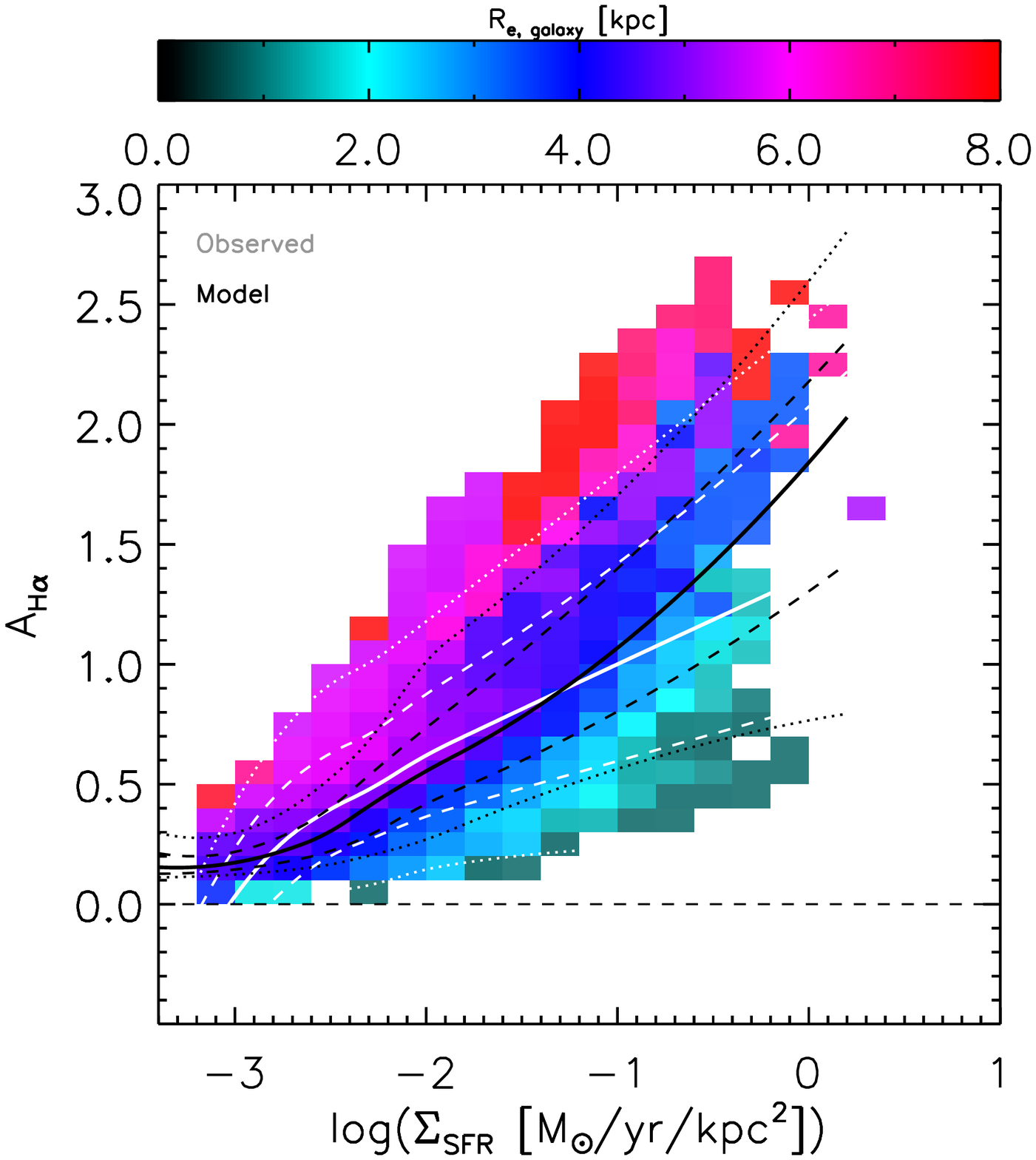}
 \includegraphics[width = .32\linewidth]{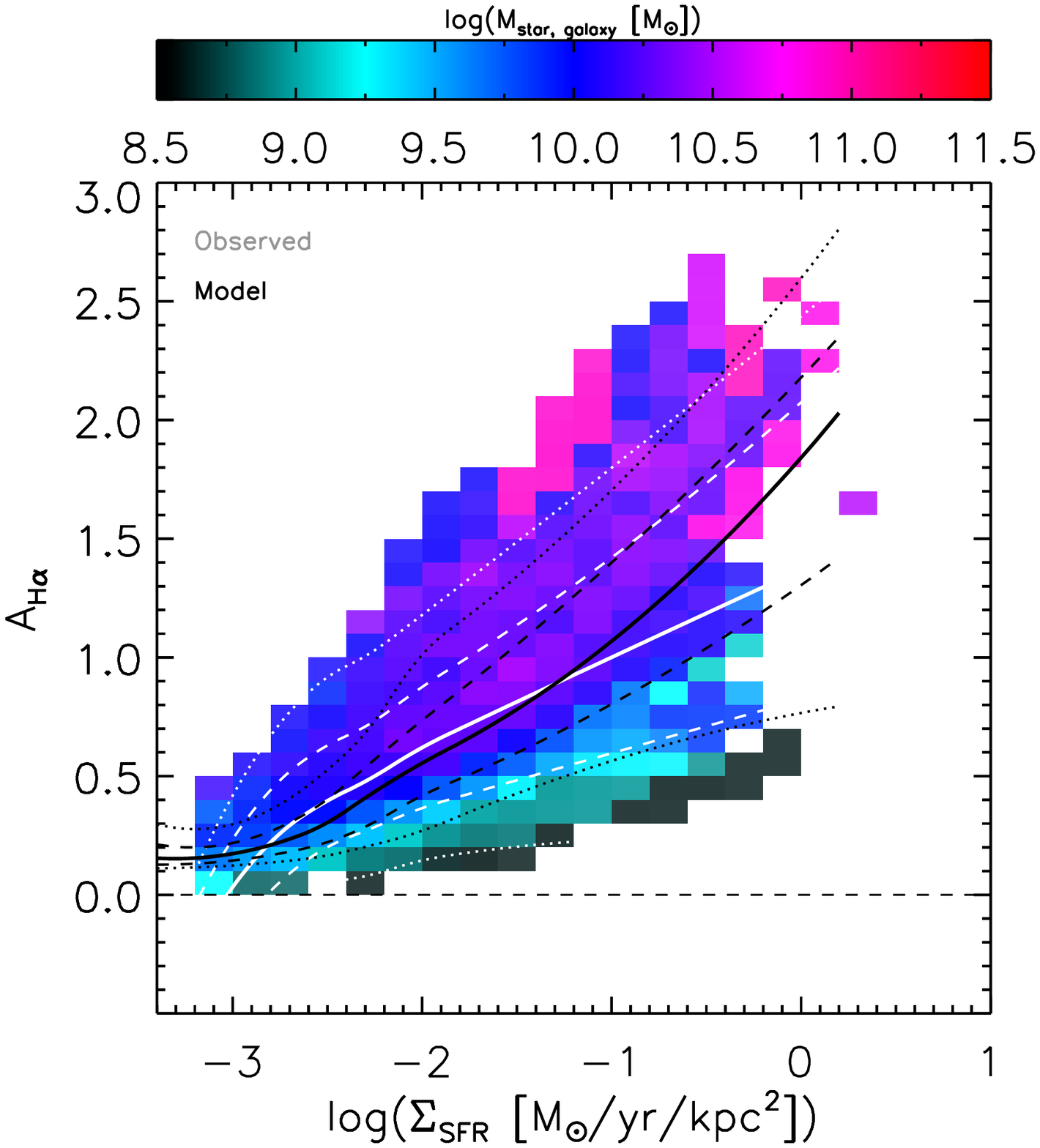}
 }
\caption{The observed star formation - extinction relation ({\it white curves}) contrasted with predictions by the same model as Figure\ \ref{without_alphabeta.fig} ({\it black curves and color coding}).  The color coding marks the median gas-phase metallicity, \Ha\ EW and galactocentric distance of the spaxels and the ellipticity, effective radius and stellar mass of their galaxies as predicted by the best-fit three-parameter model.}
\label{without_alphabeta_dependencies.fig}
\end{figure*}
%%%%%%%%%%%%%%%%%%%%% FIG 8 %%%%%%%%%%%%%%%%%%%%%
\begin{figure*}%[b]
\centering
{
 \includegraphics[width = .32\linewidth]{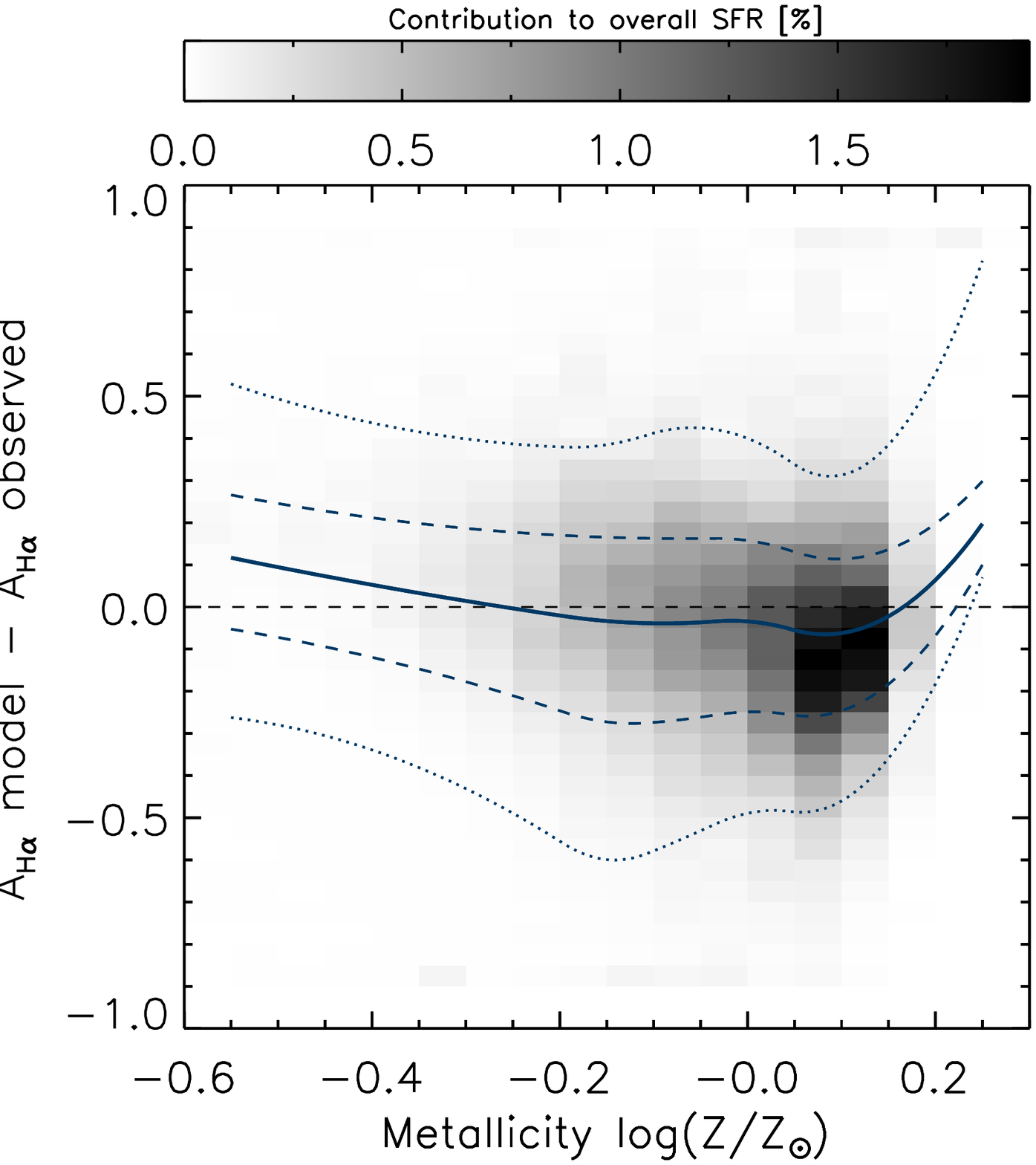}
 \includegraphics[width = .32\linewidth]{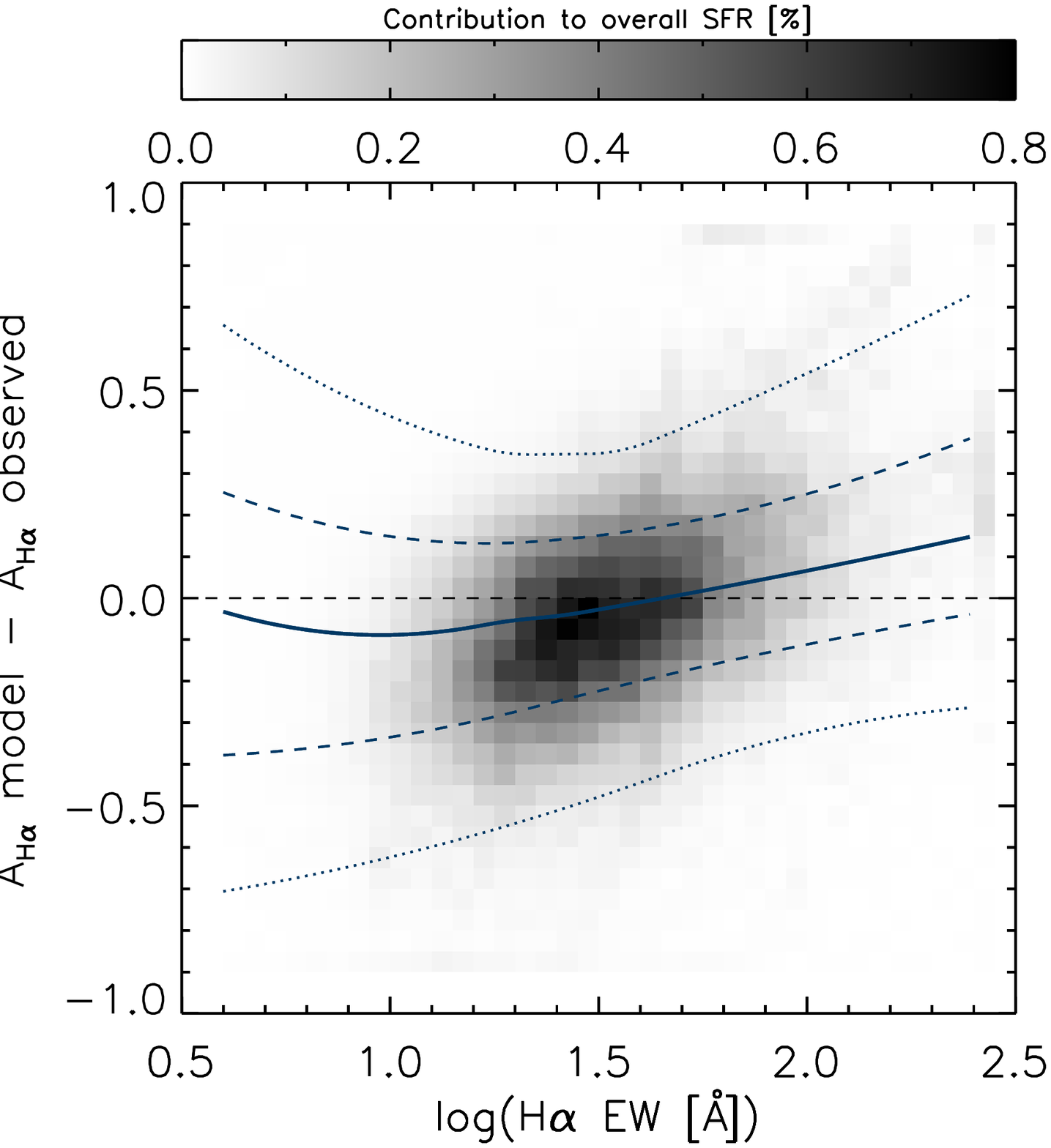}
 \includegraphics[width = .32\linewidth]{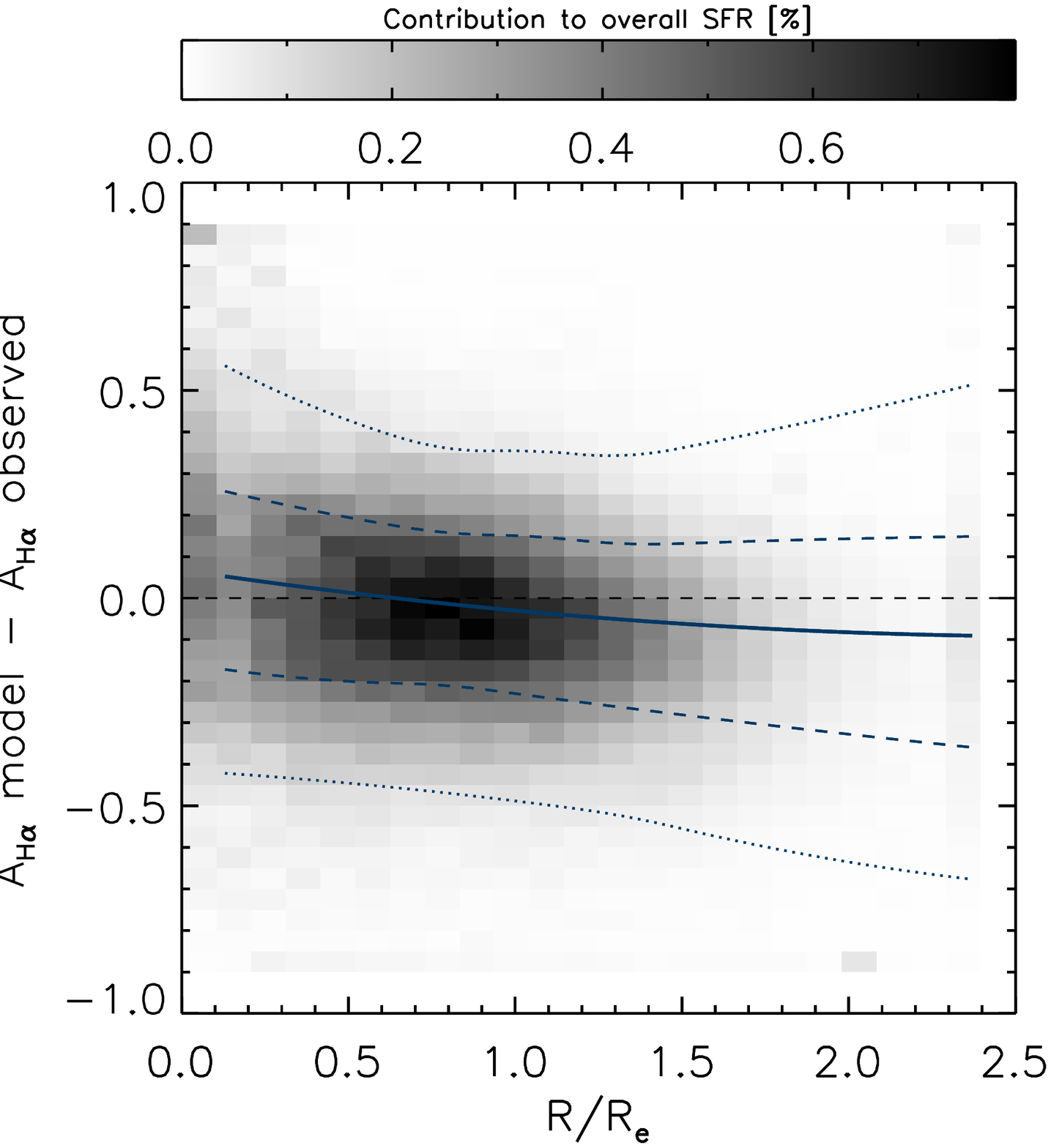}
}

{
\vspace{2mm}
 \includegraphics[width = .32\linewidth]{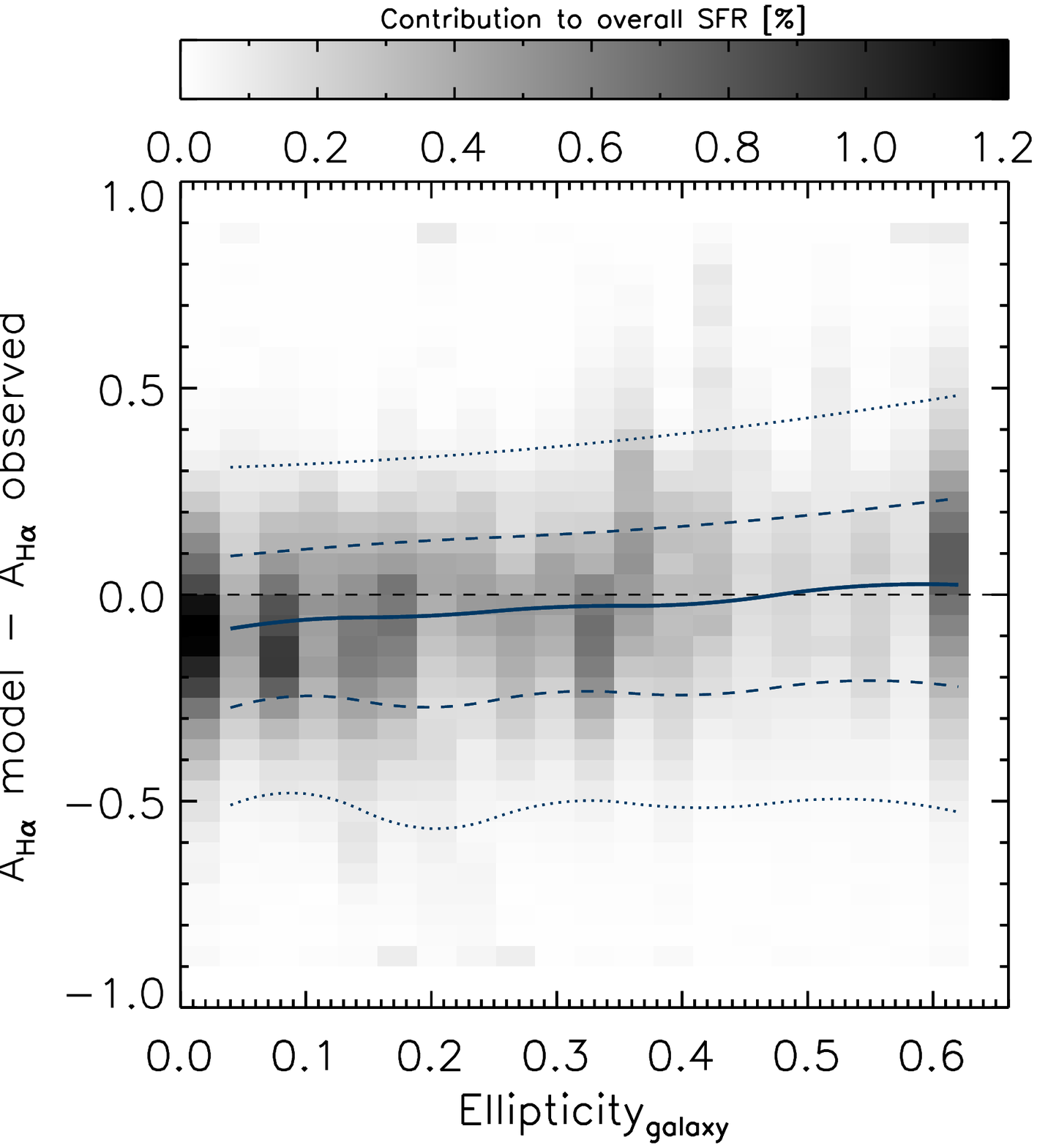}
 \includegraphics[width = .32\linewidth]{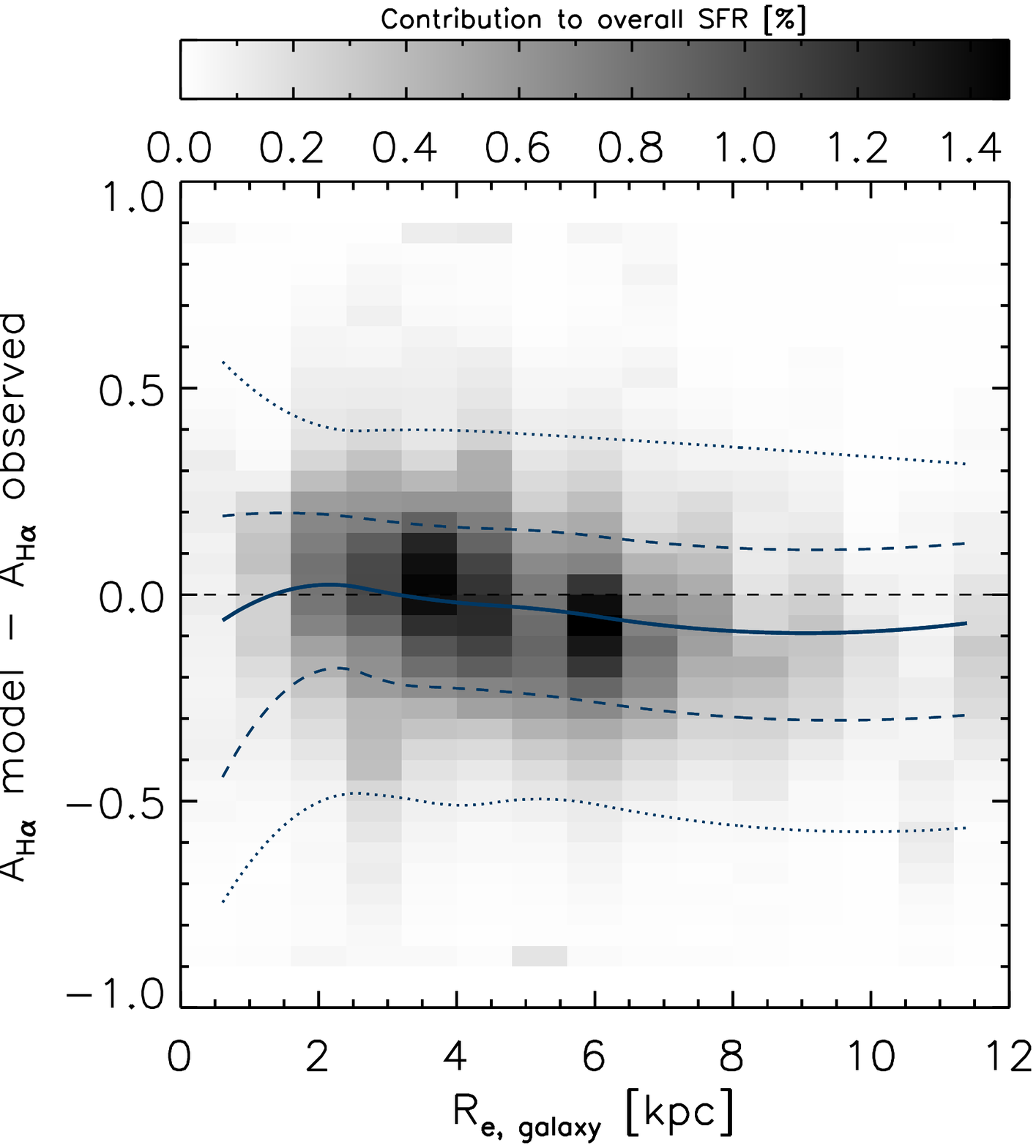}
 \includegraphics[width = .32\linewidth]{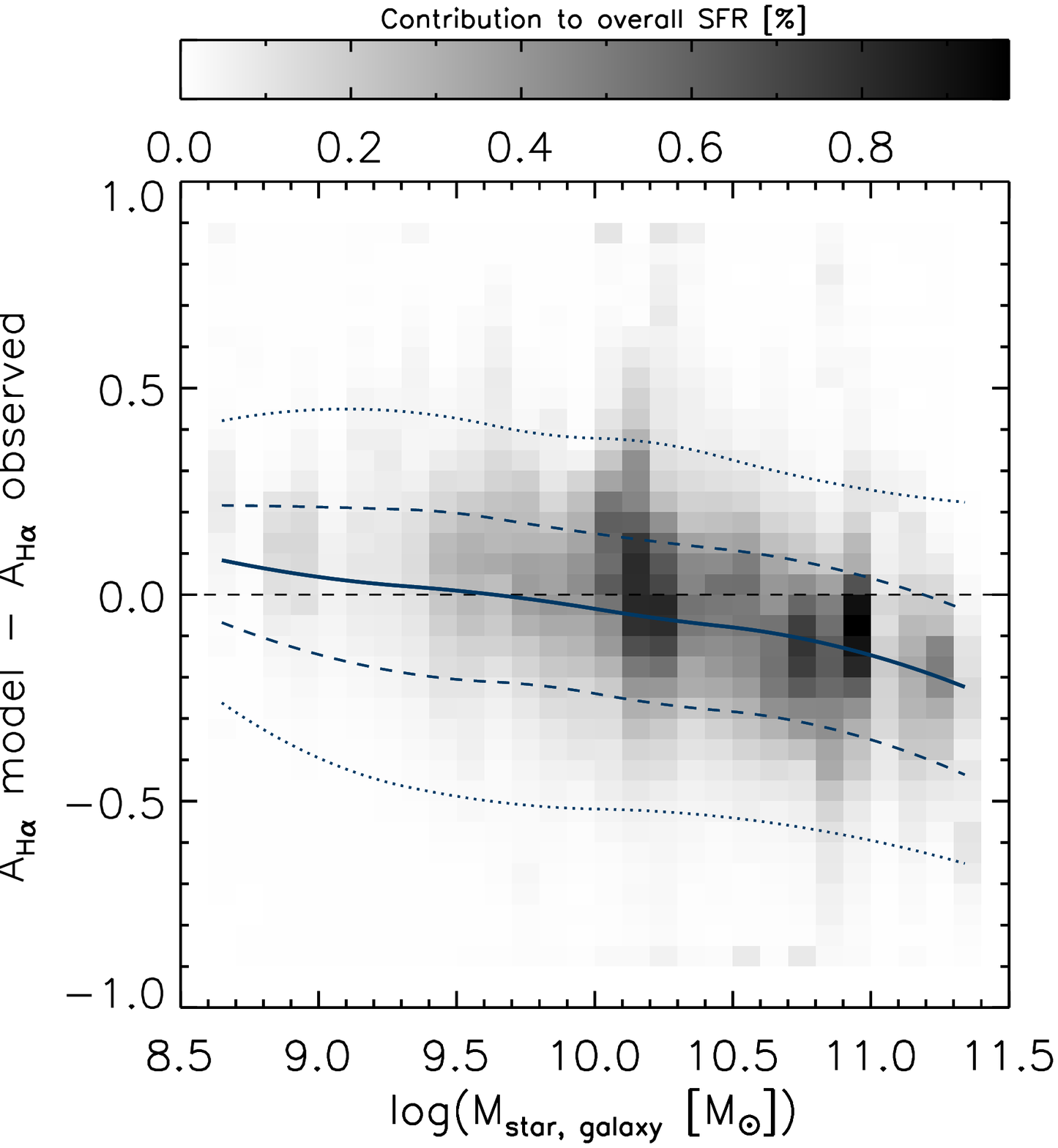}
 }
\caption{Modeled minus observed dust extinction as a function of metallicity, \Ha\ EW, galactocentric distance, and as a function of the ellipticity, size and stellar mass of the galaxies to which the spaxels belong.  Model predictions are based on the same three-parameter model presented in Figures\ \ref{without_alphabeta.fig} and\ \ref{without_alphabeta_dependencies.fig}.}
\label{without_alphabeta_residuals.fig}
\end{figure*}
}
%%%%%%%%%%%%%%%%%%%%%%%%%%%%%%%%%%%%%%%%%%%%

Thus far, we forced the KS slope to remain fixed at unity ($n_{KS} = 1$) when fitting a model. Alleviating this constraint (i.e., leaving $n_{KS}$ free) markedly improves the match between model predictions and observations.  Figures\ \ref{without_alphabeta.fig} to\ \ref{without_alphabeta_residuals.fig} present the predictions of a model with free KS slope, $f_{screen}$ and $N_{clumps}$.  In other words, we have still kept the parameters $\alpha$ and $\beta$ from Equation\ \ref{fscreen.eq} fixed to 0, meaning that the relative breakdown between dust in a mixture and foreground component is identical for all spaxels of all galaxies.  The improvement of this model compared to those discussed in Section\ \ref{fixed_models.sec} can be evaluated in terms of the reduced chi-squared value of the fit ($\chi^2_{red} = 2.81$), but is especially also apparent from the good correspondence over the full dynamic range of $\Sigma_{SFR}$, not just where most of the star formation is contributed (Figure\ \ref{without_alphabeta.fig}).  The improved match to the observed trends can further be appreciated from Figure\ \ref{without_alphabeta_dependencies.fig} which visualizes the secondary dependencies (i.e., the model equivalent of Figure\ \ref{dependencies.fig}) and Figure\ \ref{without_alphabeta_residuals.fig} which shows the modeled minus observed $A_{\eqHa}$ residuals.

Nevertheless, the model still exhibits a few shortcomings. It slightly underestimates the scatter in extinction values at intermediate $\Sigma_{SFR}$ and overpredicts the median $A_{\eqHa}$ at the highest, albeit sparsely populated $\Sigma_{SFR}$. Regarding secondary dependencies, the model reproduces many features in a qualitative manner at least: outliers towards higher extinction are more typically characterized by higher gas-phase metallicities, lower \Ha\ EWs, higher ellipticities and larger galaxy sizes and masses. Conversely, spaxels lying below the median relation are relatively metal-poor, feature higher \Ha\ EWs, lower ellipticities and smaller galaxy sizes and masses. This is to a large degree the case by construction of course, although we note that at this stage neither \Ha\ EW measurements of the spaxels nor stellar masses of the galaxies have been used as input, the way inclination enters the model does not involve any free parameters, and the only freedom in metallicity-dependent conversion steps has been the small $-2.3 < \log(DGR(Z_{\odot})) < -2.0$ interval dictated by independent observational results (see Section\ \ref{DGR.sec} and references therein). The existence of a relatively small number of spaxels at low $\Sigma_{SFR}$ yet high $A_{\eqHa}$ corresponding to regions at large galactocentric distance ($R/R_e$) is not captured by this 3-parameter model where we have still kept $\alpha = 0$ and $\beta = 0$ in Equation\ \ref{fscreen.eq}. Figure\ \ref{without_alphabeta_residuals.fig} further shows that residuals exhibit minor systematic trends with \Ha\ EW, $R/R_e$ and galaxy stellar mass, suggesting in the former case that while a relation between offset from the median star formation -- extinction relation and \Ha\ EW emerges naturally, the extent to which high and low EW spaxels are differentiated in the predicted $A_{\eqHa}$ is still underestimated. Finally, and arguably of most concern, the best-fit KS slope is highly super-linear at $n_{KS} = 2.12$, significantly above slopes quoted in the literature for the molecular gas -- star formation law.

%%%%%
% FIG 9
%%%%%
\begin {figure}[t]
%\epsscale{0.49}
\plotone{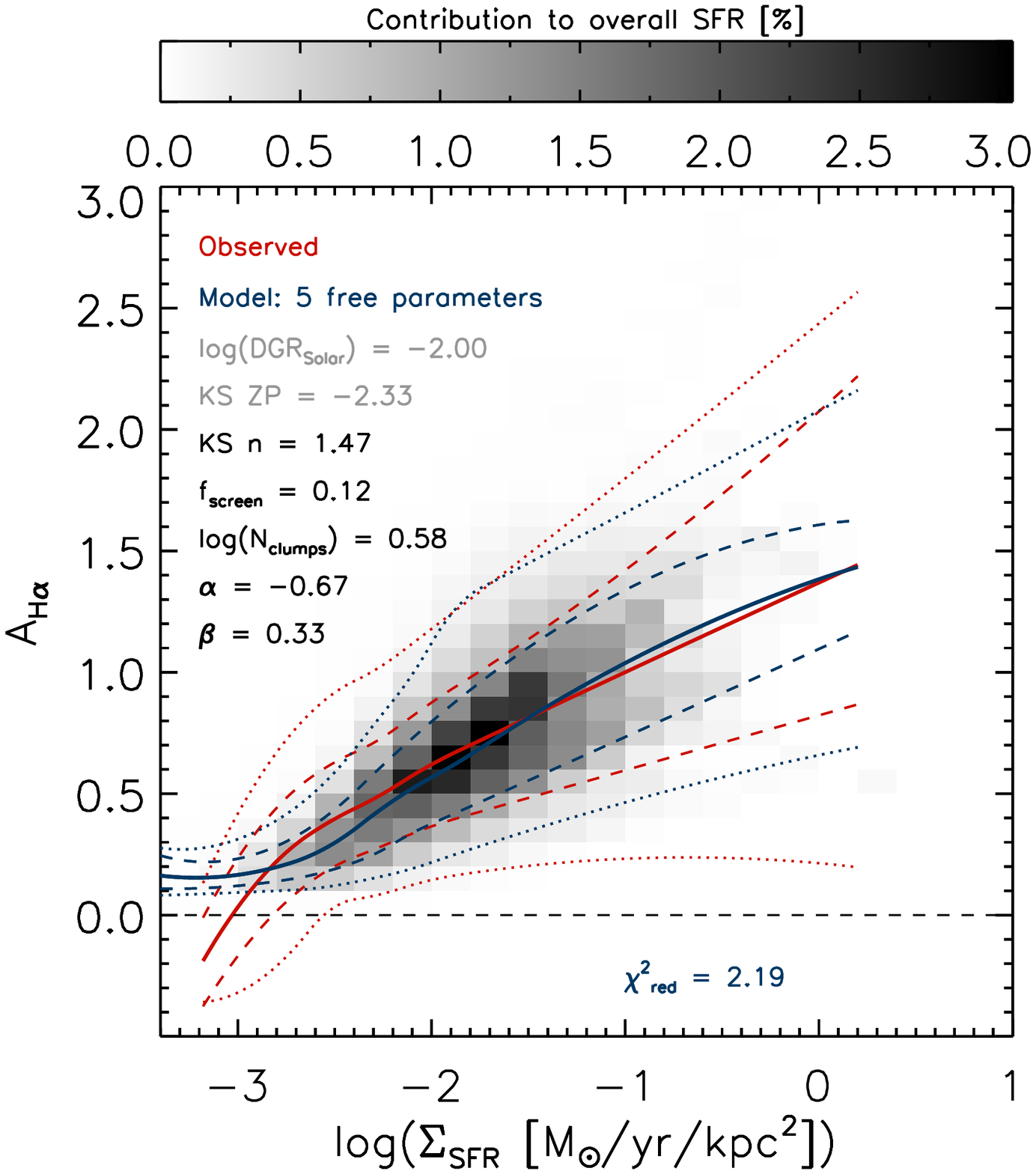}
\caption{The observed star formation -- extinction relation ({\it red curves}) contrasted with our favored best-fit five-parameter model allowing a range of dust geometries and free KS slope ({\it blue curves and grayscales}). Listed are best-fit values of $f_{screen}$, $\log(N_{clumps})$, KS relation slope ($n_{KS}$), $\alpha$ and $\beta$. }
\label{5free.fig}
%\vspace{0.5cm}
\end{figure}

%%%%%
% FIG 10
%%%%%
\begin{figure*}
\centering
{
 \includegraphics[width = .32\linewidth]{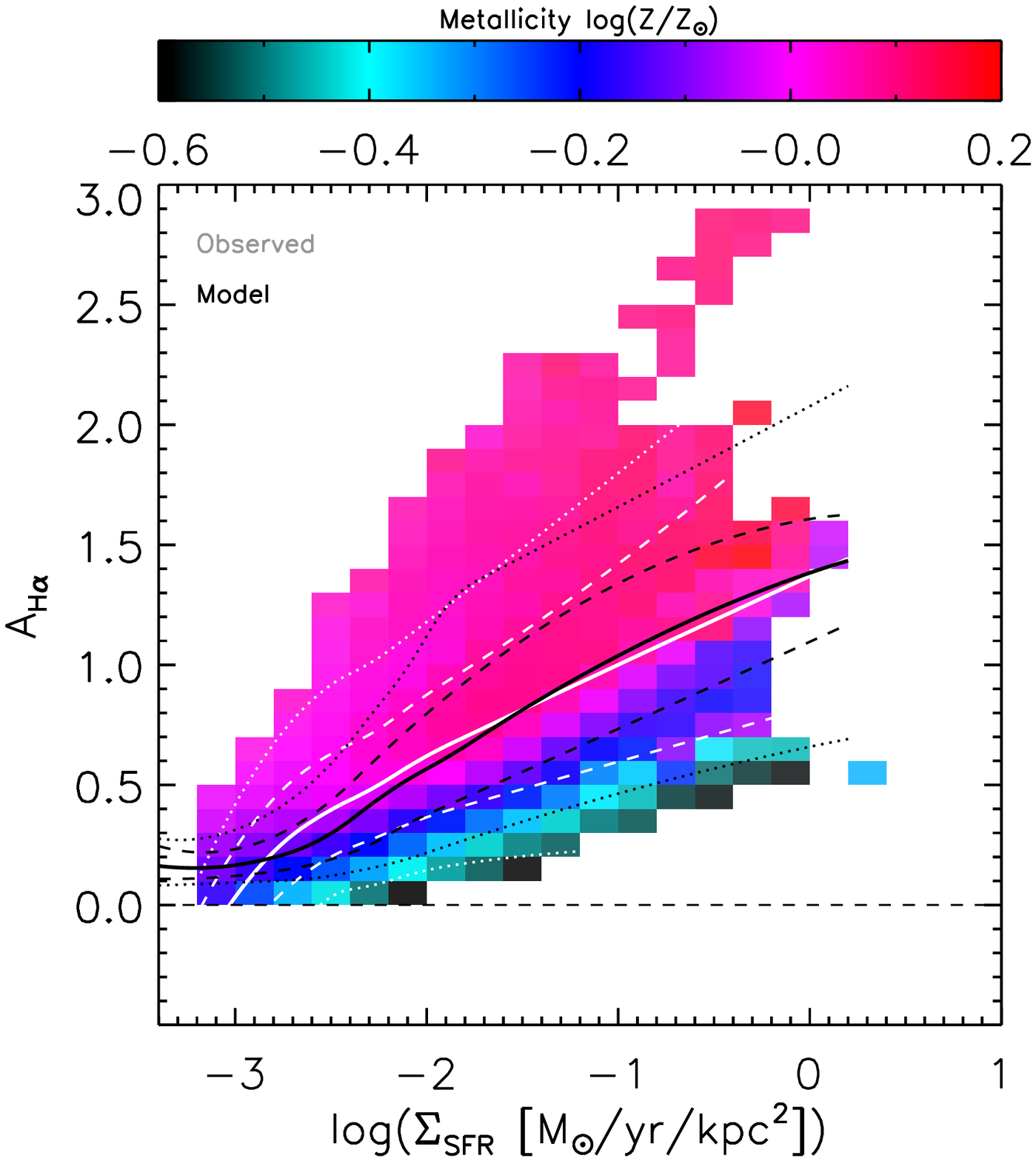}
 \includegraphics[width = .32\linewidth]{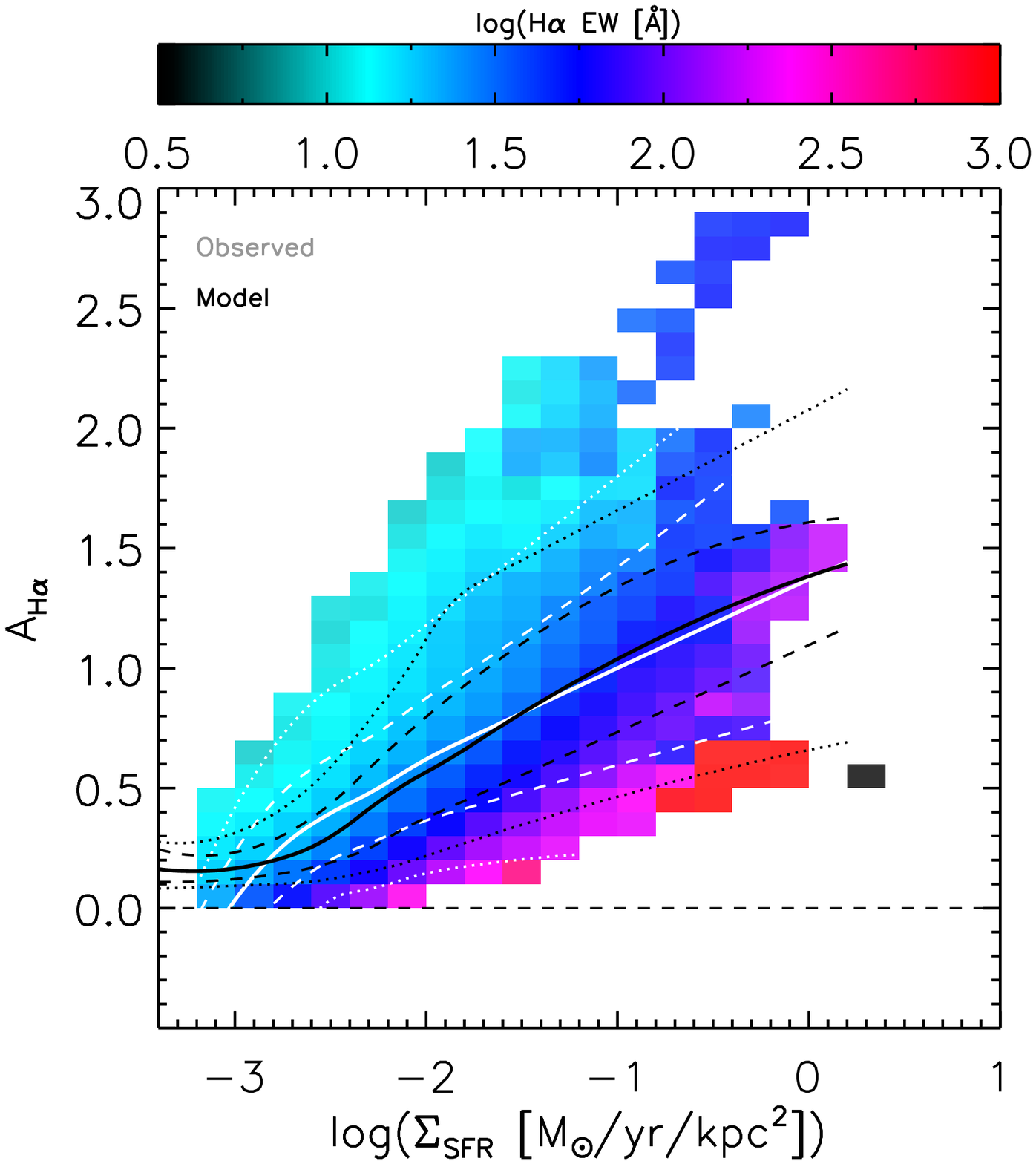}
  \includegraphics[width = .32\linewidth]{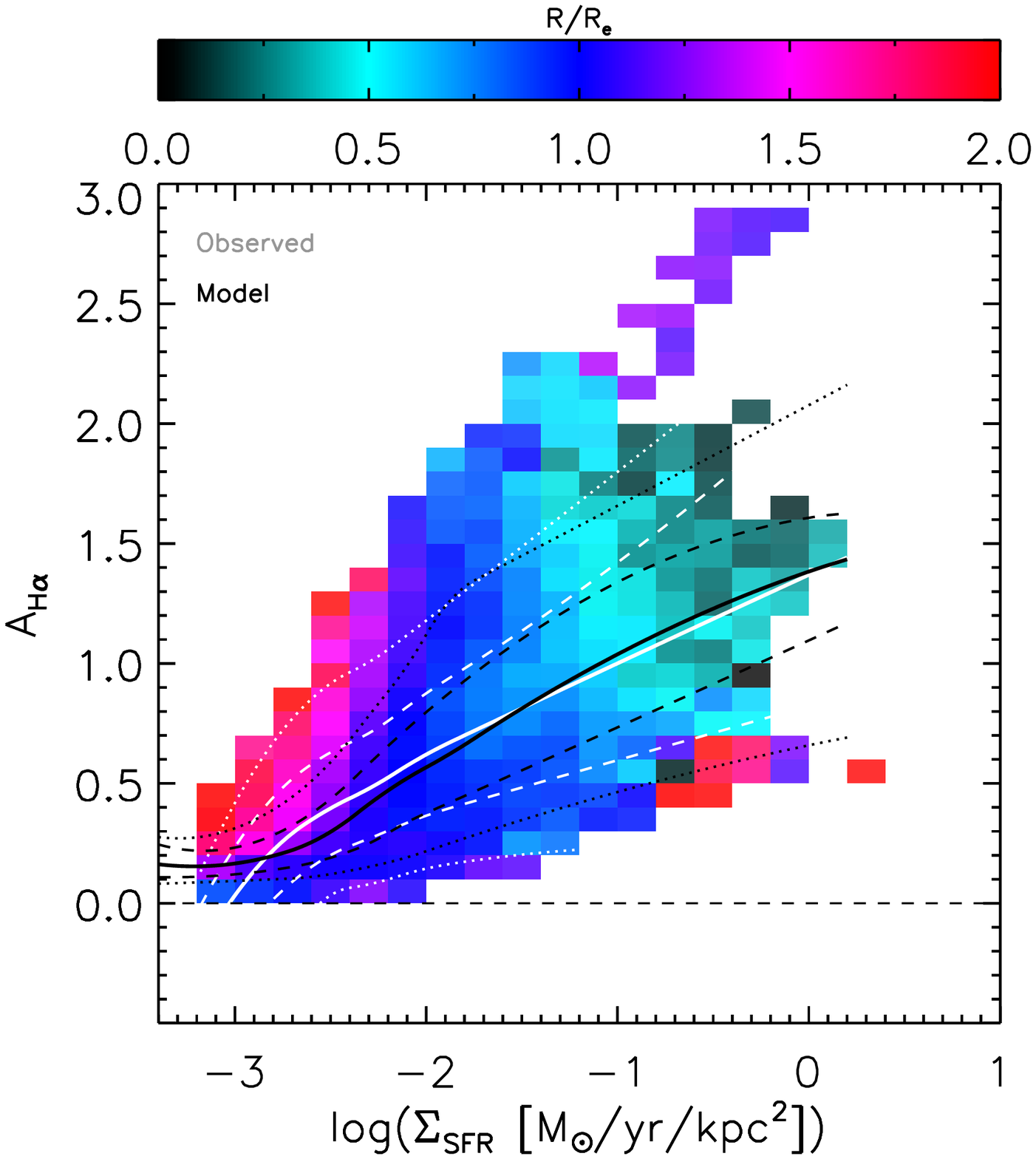}
}

{
 \vspace{2mm}
 \includegraphics[width = .32\linewidth]{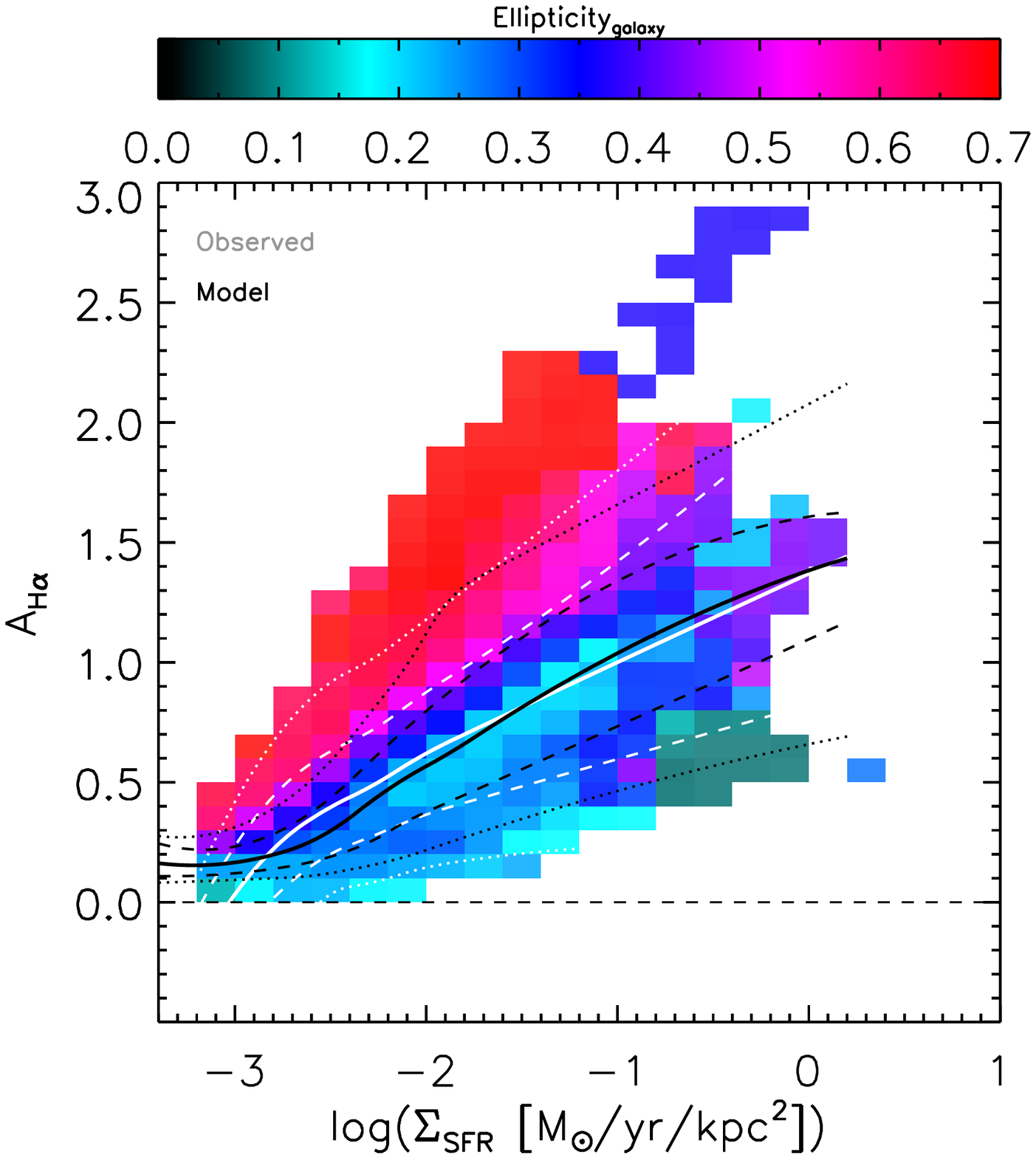}
 \includegraphics[width = .32\linewidth]{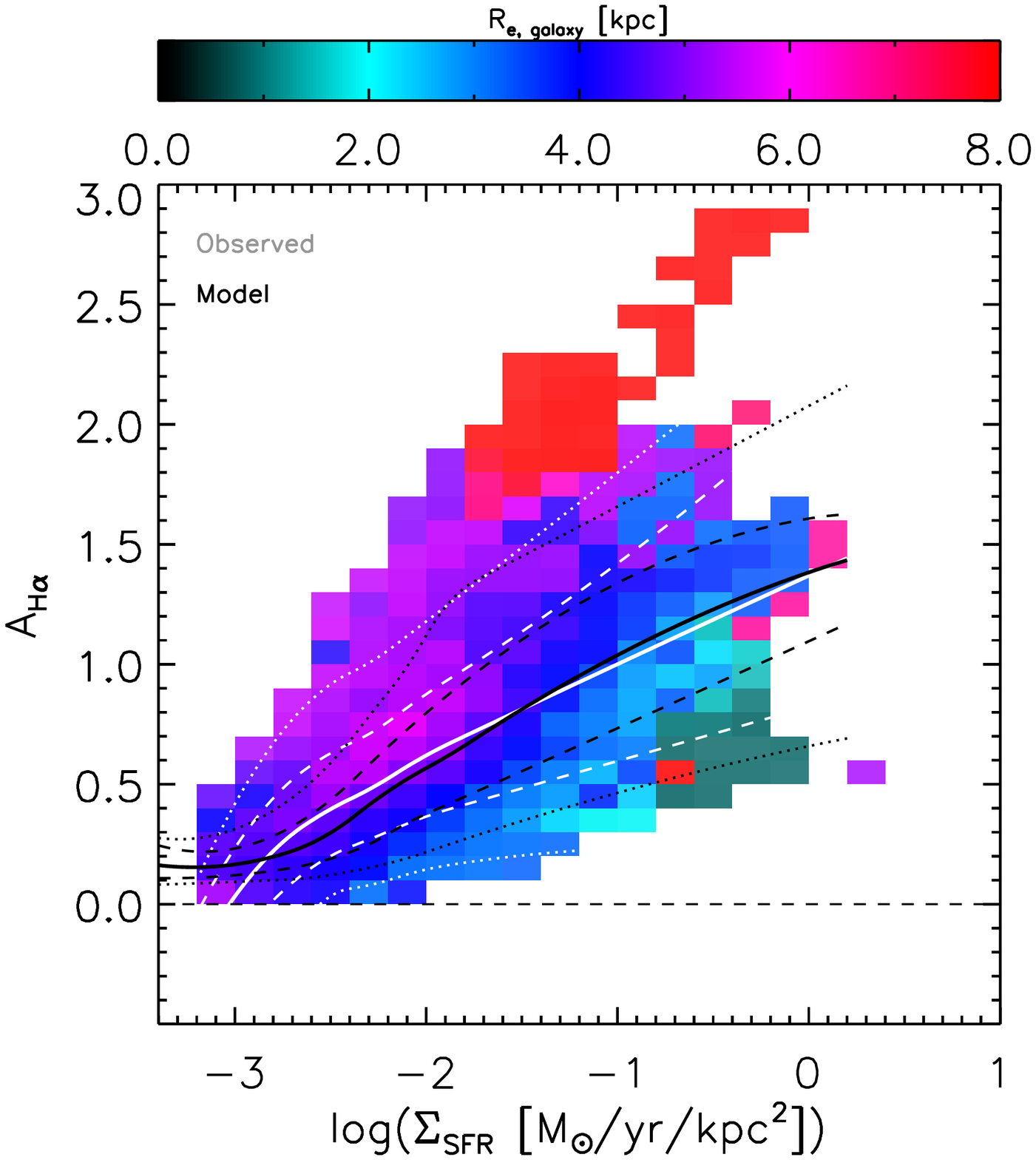}
 \includegraphics[width = .32\linewidth]{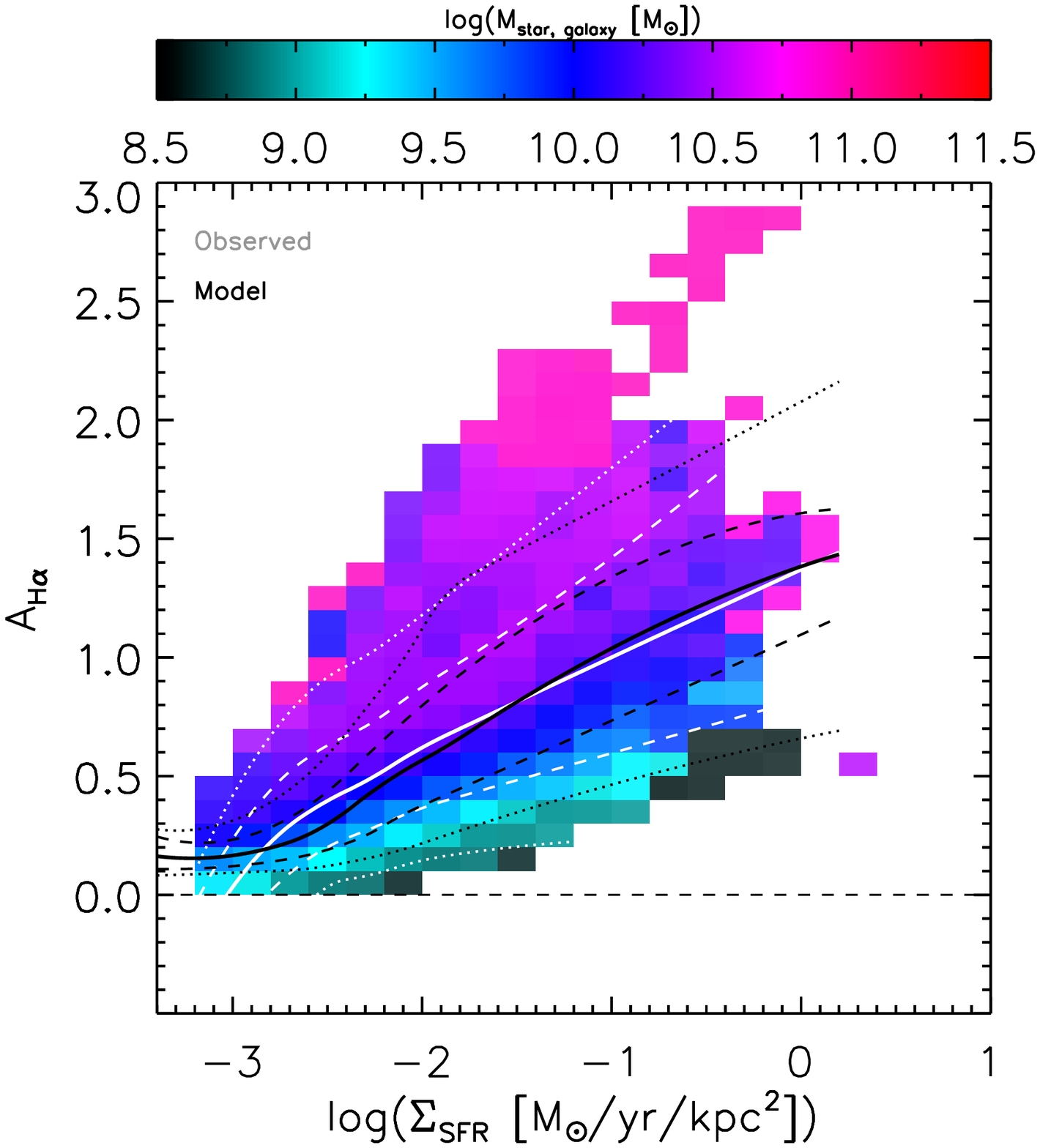}
 }
\caption{ The observed star formation -- extinction relation ({\it white curves}) contrasted with the model relation ({\it black curves}) color-coded by the median gas-phase metallicity, \Ha\ EW and galactocentric distance of spaxels in the bins, and ellipticity, effective radius and stellar mass of their galaxies as predicted by our favored model.}
\label{5free_dependencies.fig}
%\vspace{0.5cm}
\end{figure*}

%%%%%
% FIG 11
%%%%%
\begin{figure*}
\centering
{
 \includegraphics[width = .32\linewidth]{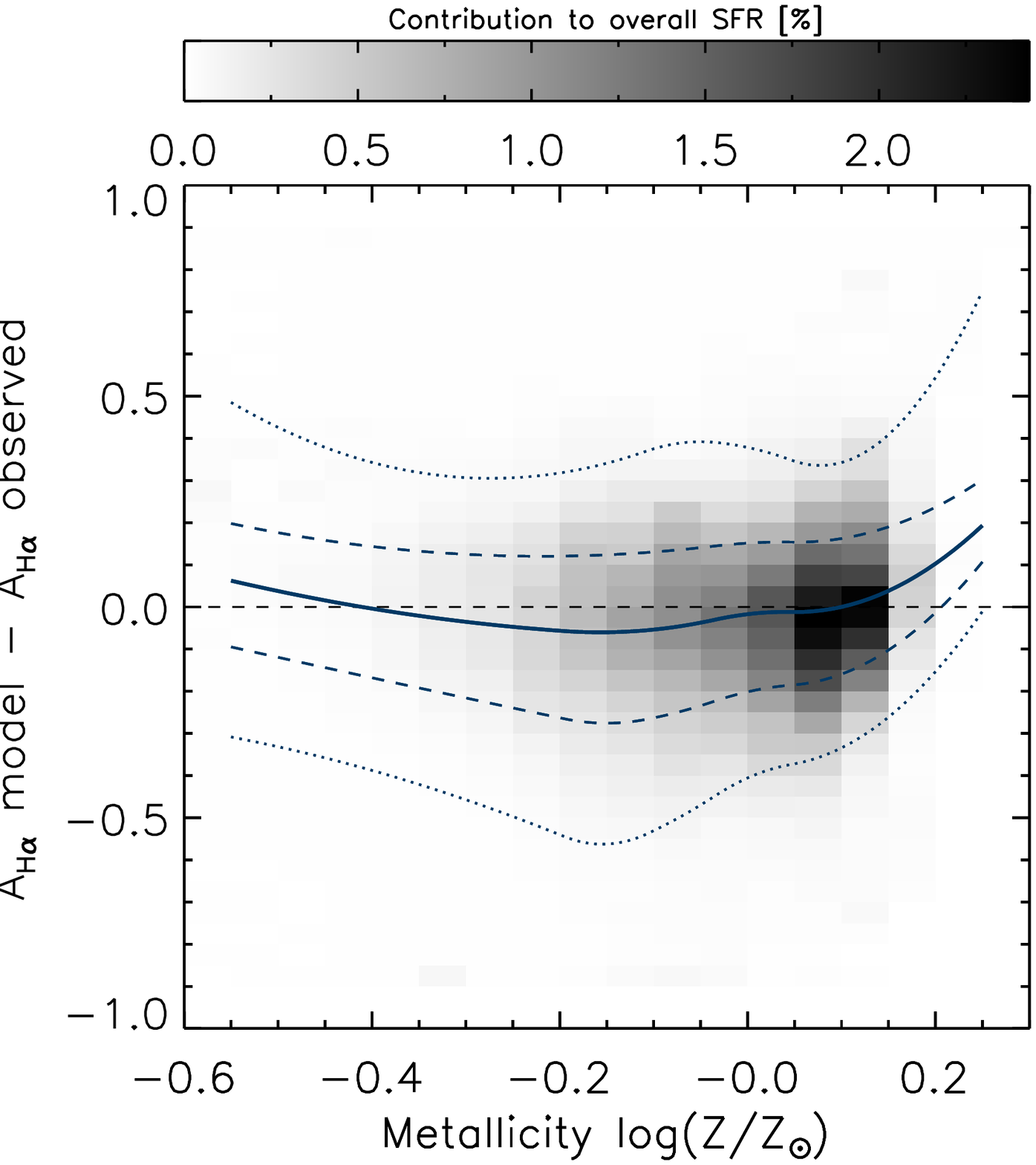}
 \includegraphics[width = .32\linewidth]{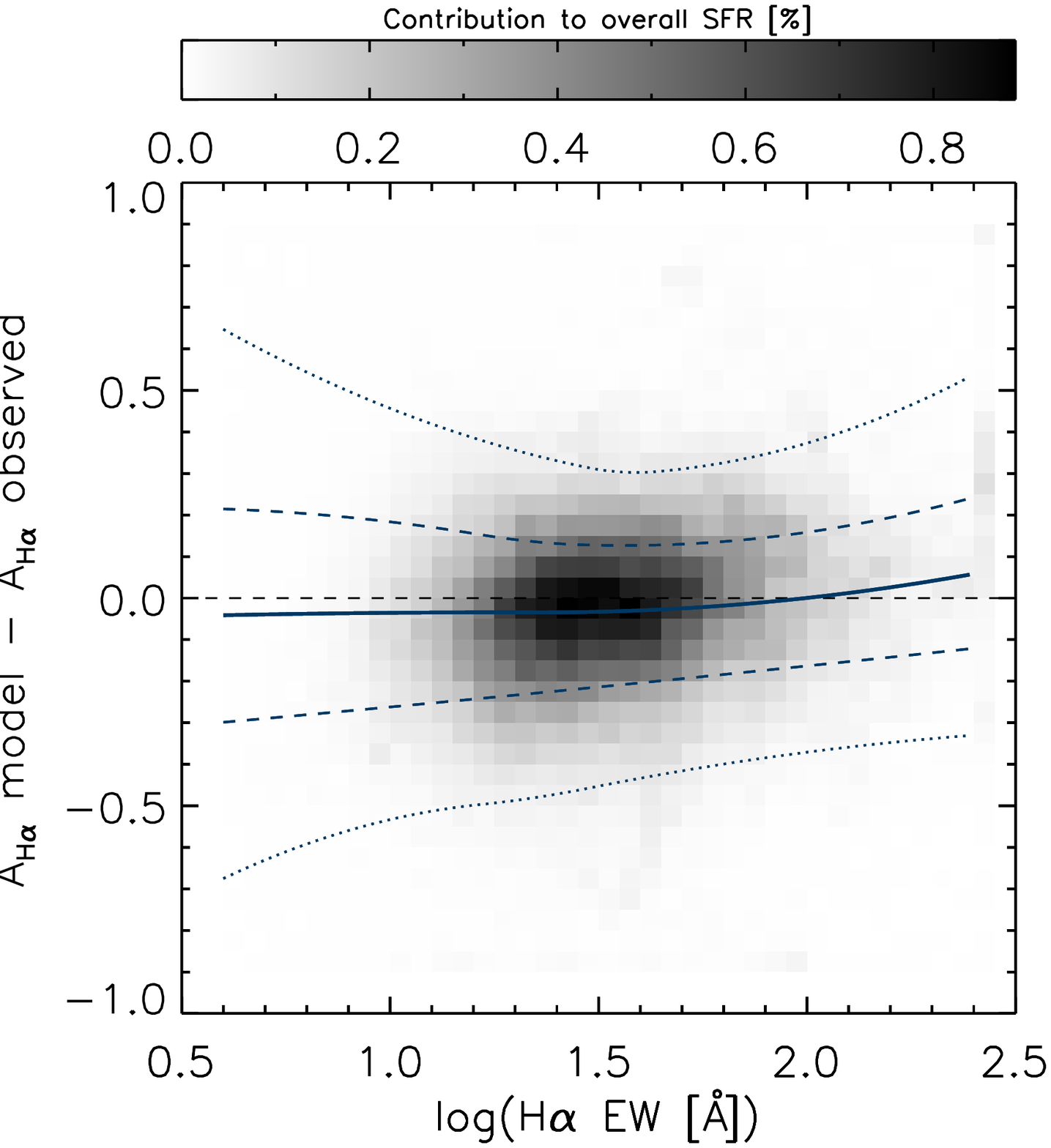}
 \includegraphics[width = .32\linewidth]{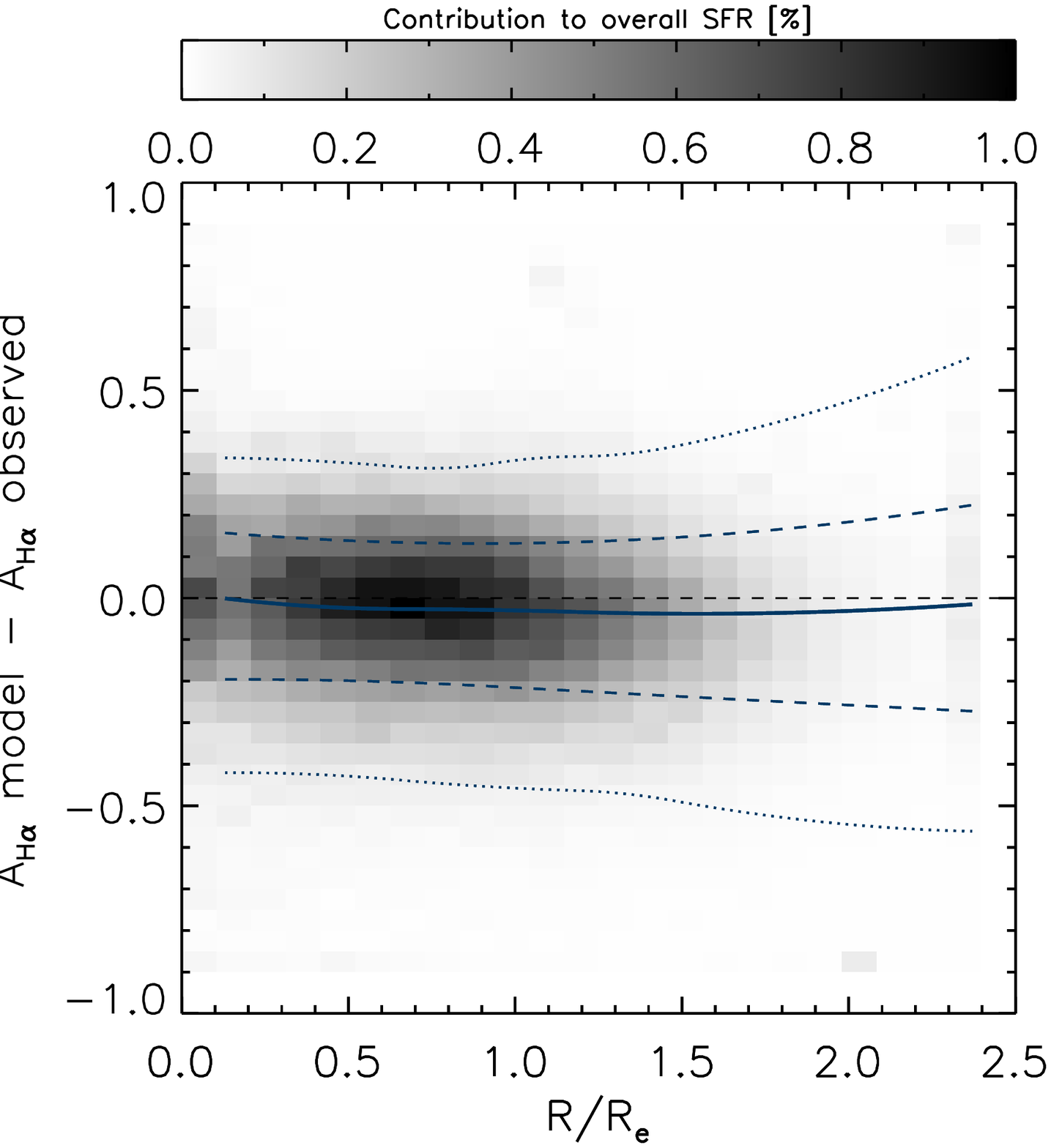}
}

{
\vspace{2mm}
 \includegraphics[width = .32\linewidth]{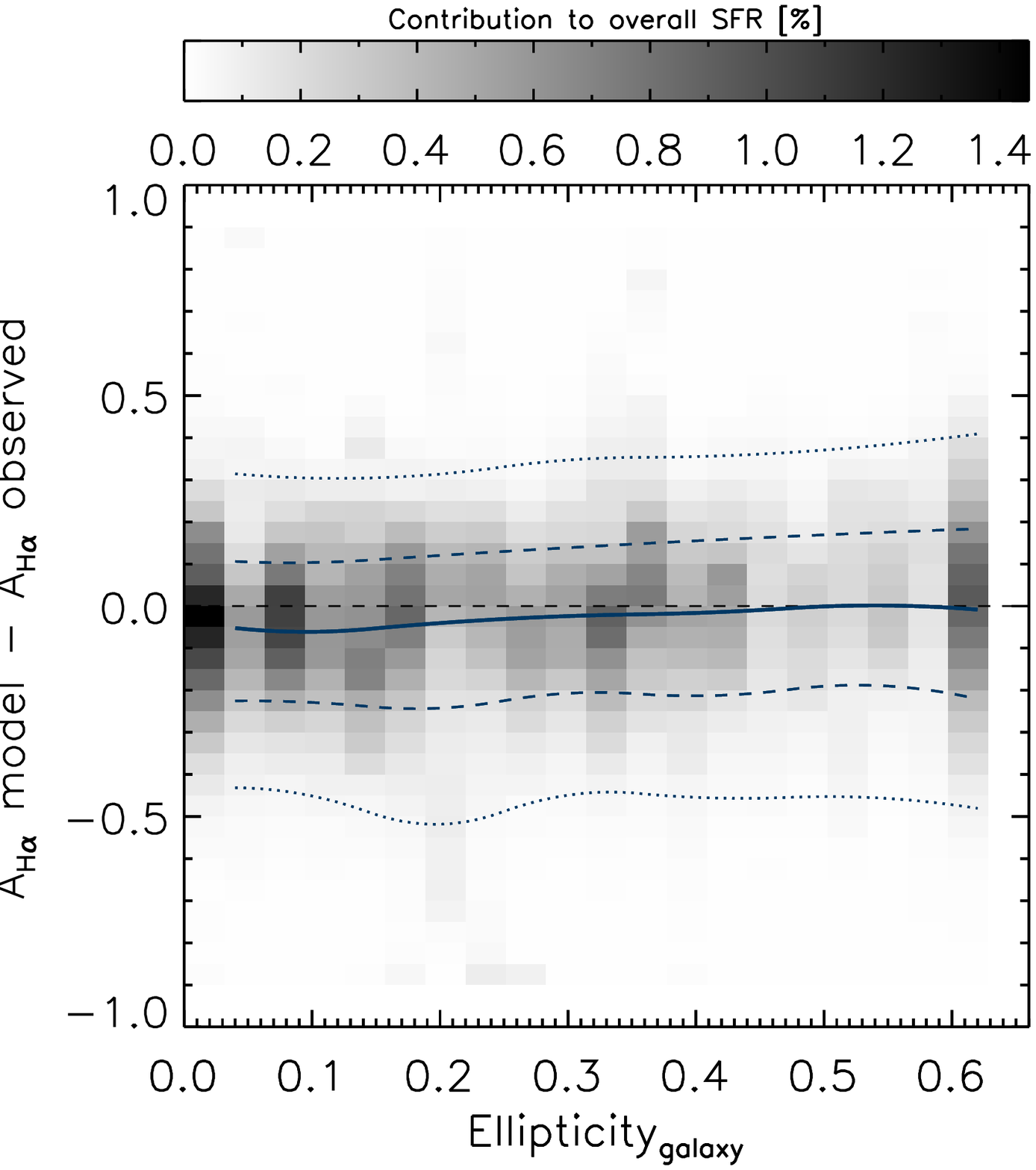}
 \includegraphics[width = .32\linewidth]{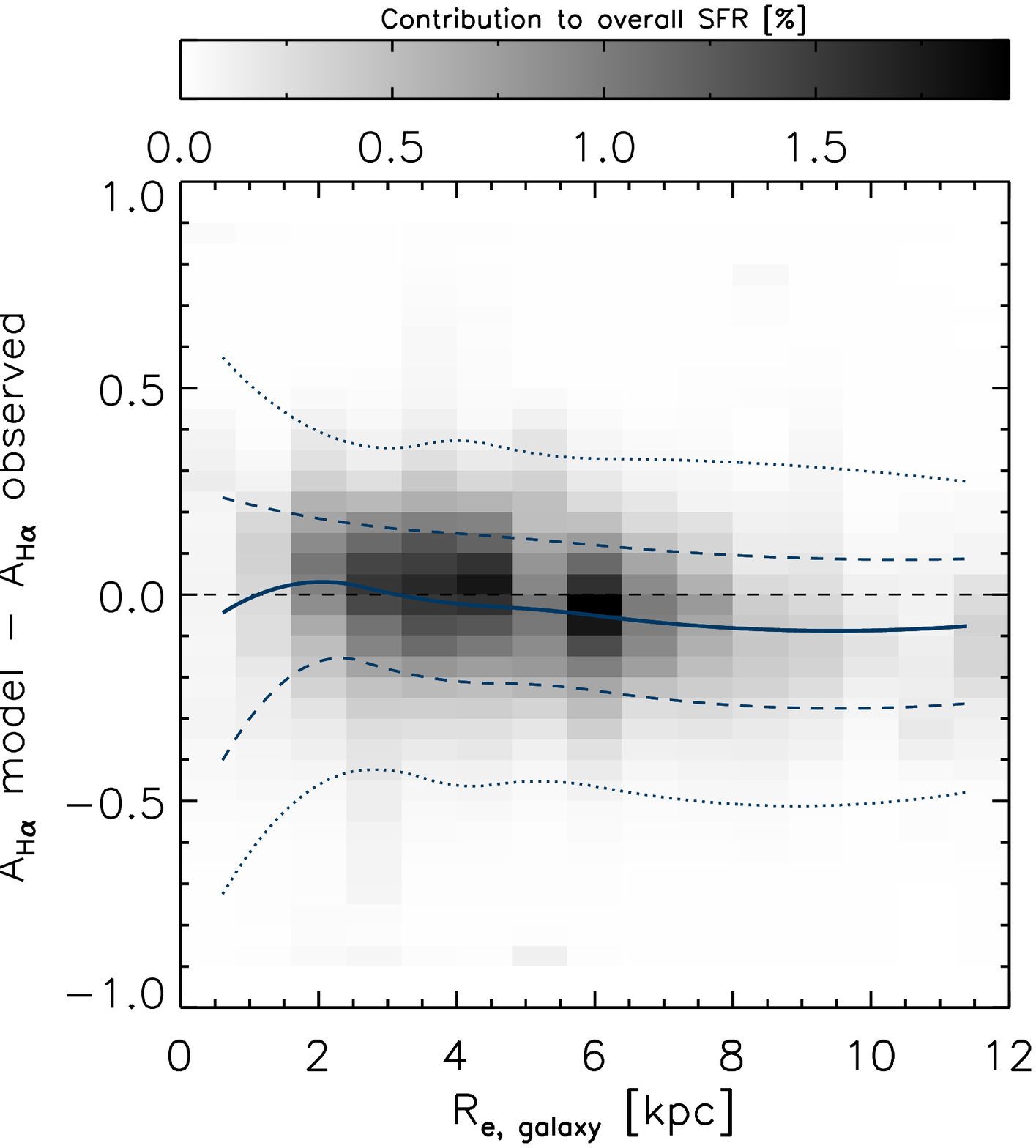}
 \includegraphics[width = .32\linewidth]{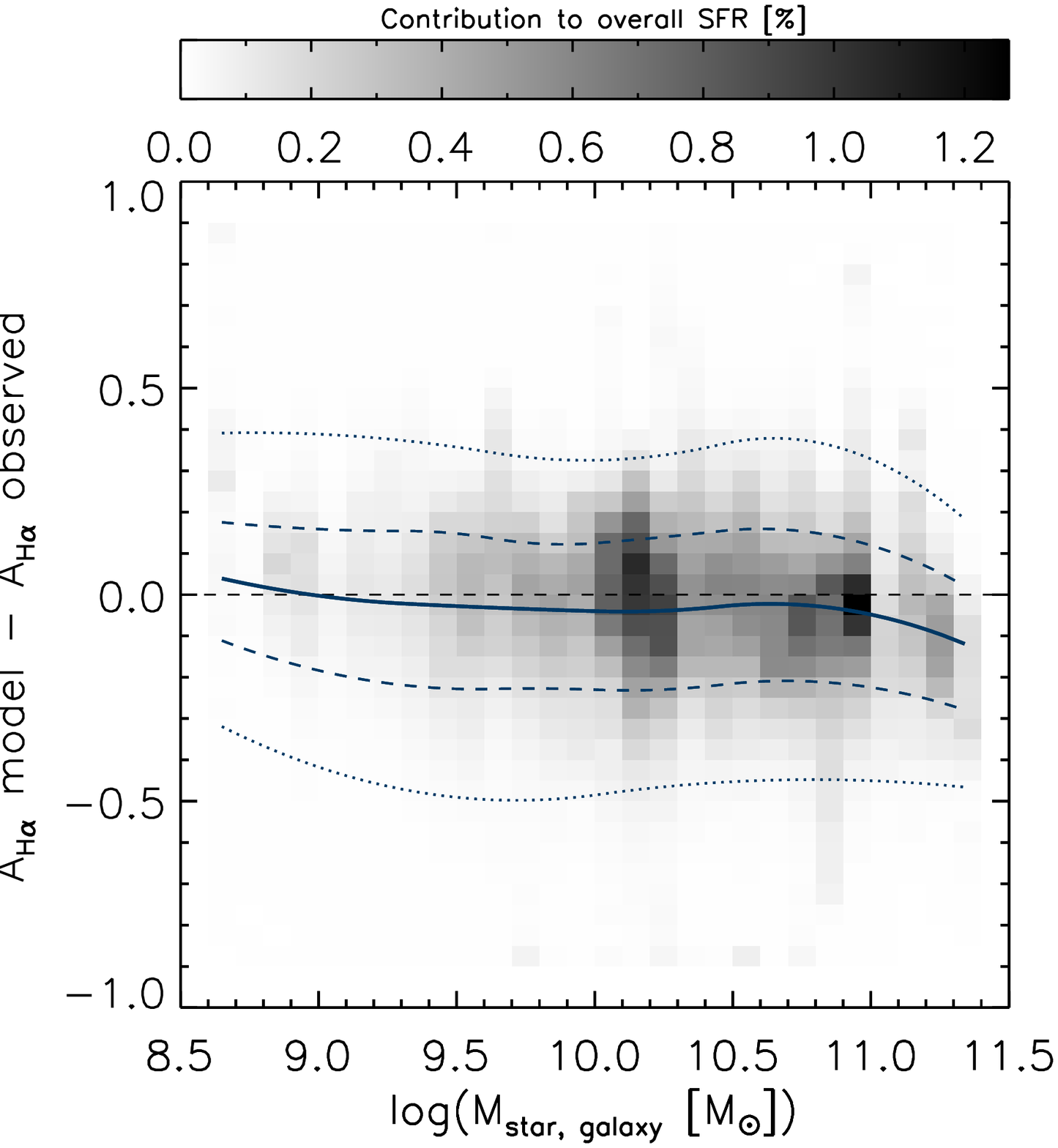}
 }
\caption{Modeled minus observed dust extinction for our favored five-parameter model plotted versus physical properties of the spaxels and of the galaxies the spaxels belong to. Blue curves represent the median residual (plus 68th and 95th percentiles) of spaxels in a given bin. Grayscales indicate the relative contributions to the summed SFR over all spaxels. No systematic trends of residuals with the considered observables are evident.}
\label{5free_residuals.fig}
%\vspace{0.5cm}
\end{figure*}

The above considerations prompt us to introduce the \Ha\ EW and galactocentric distance dependence of $f_{screen}$ as parameterized in Equation\ \ref{fscreen.eq}. This now represents our full and favored model, which improves upon the shortcomings of the 3-parameter model described above. In other words, the $\alpha$ and $\beta$ parameters are now left free, such that the fraction of dust in a foreground screen is no longer the same for all spaxels. This is physically plausible as the spaxels in our sample are drawn from galaxies with a wide range in properties, spanning for example 3 orders of magnitude in stellar mass, and within them range from very central locations to the outskirts at $\sim 2 R_e$. We present the resulting star formation - extinction relation in Figure\ \ref{5free.fig}, its secondary dependencies in Figure\ \ref{5free_dependencies.fig}, and the residuals in Figure\ \ref{5free_residuals.fig}.

%...above.  \footnote{For completeness, we present the intermediate model, with $\alpha$ free and $\beta = 0$ in the Appendix Figures\ \ref{without_beta.fig} to\ \ref{without_beta_residuals.fig}.}

The star formation - extinction relation, including its scatter, is reasonably well reproduced over the full $\Sigma_{SFR}$ range. The tail of (low $\Sigma_{SFR}$; high $A_{\eqHa}$) spaxels corresponding to locations in galaxy outskirts is reproduced since we adopted a larger $f_{screen}$ for them (Equation\ \ref{fscreen.eq}) and with more of the dust in a foreground component the extinction for a given dust column becomes more effective. Introducing $\alpha$ and $\beta$ flattened the systematic residuals with respect to \Ha\ EW and $R/R_e$. The scatter in residuals $A_{\eqHa\ model} - A_{\eqHa\ observed}$ is of order $\sim 0.2$ mag. The lack of systematic trends in the residual plots (Figure\ \ref{5free_residuals.fig}) implies that there is no clear evidence prompting the addition of further complexity/parameters to the model.

We derive uncertainties on the best-fit model parameters by bootstrapping the sample of galaxies from which spaxels are drawn.  In other words, similar to how we derived uncertainties for the best-fit (empirical) linear relation described by Equation\ \ref{linear_fit.eq}, we create 100 mock samples by randomly resampling the galaxies with replacement.  We next fit our model to the ensemble of spaxels for each of the 100 mock samples and compute central 68th percentiles of the parameter values, yielding confidence intervals of $n_{KS} = 1.47 \pm 0.04$, $f_{screen} = 0.12 \pm 0.01$, $\log(N_{clumps}) = 0.58 \pm 0.07$, $\alpha = -0.67 \pm 0.05$ and $\beta = 0.33 \pm 0.02$.

Importantly, while still significantly super-linear at $n_{KS} = 1.47$ the KS slope no longer exceeds the range of values quoted in the literature based on CO observations. Having introduced additional freedom in our model compared to the simple models explored in Figure\ \ref{naive_models.fig}, we checked that forcing $n_{KS} = 1$ while leaving $\alpha$ and $\beta$ free does not yield an equally satisfactory agreement between model and observations. The reduced chi-squared, driven primarily by the spaxels contributing the bulk of the overall SFR, increases modestly to $\chi_{red}^2 = 2.51$. The median star formation -- extinction relation predicted by such a model with $n_{KS} = 1$ differs in shape from the near-linear relation found empirically, underestimating the median $A_{\eqHa}$ for $-2.8 < \log(\Sigma_{SFR}) < -2$ and more dramatically so at $\log(\Sigma_{SFR}) > -0.9$. In addition, more pronounced systematic trends are observed between the $A_{\eqHa\ model} - A_{\eqHa\ observed}$ residuals and other observables, most notably metallicity and \Ha\ EW. The flexibility to allow for a non-linear KS relation thus seems desired to explain the empirically established $A_{\eqHa} - \Sigma_{SFR}$ relation. This remains the case if different assumptions are made regarding scattering of light in the foreground screen. We remind the reader that by default we assume the foreground dust to be sufficiently far away from the emitting sources that scattering in the screen is negligible. None of the alternatives (isotropic, an-isotropic or forward-only scattering in the screen) yield model fits of better quality than our favored model, with notably the extreme and most unrealistic case of forward-only scattering yielding a poor fit. Nevertheless, we note that all such model realizations with different scattering behavior yield KS slopes in the range $1.4 < n_{KS} < 1.8$.

We point out that the match to the observations is not perfect, as can be seen for example in the secondary dependence on metallicity (top left panels of Figure\ \ref{5free_dependencies.fig} versus\ \ref{dependencies.fig}). The observations show the presence of a small number of spaxels at low $\Sigma_{SFR}$ (lying outside the 95th percentile range) with relatively high extinction yet relatively low, sub-Solar metallicities. The $R/R_e$ panel of Figure\ \ref{dependencies.fig} indicates the respective spaxels come predominantly from the outskirts of galaxies. Their location at large galactocentric distance may suggest that here the exact prescription for HI and to which degree HI dominated regions are associated with dust may play a role. We explored prescriptions for HI alternative to the default approach outlined in Section\ \ref{DGR.sec} in order to account for this small number of spaxels (less than 2\% of the overall sample) with low metallicity yet significant extinction, but found all of them to result in a poorer match for the bulk of the spaxels.

The best-fit $\alpha$ in our model is negative, implying that spaxels with high \Ha\ EW have a lower fraction of their dust column in a foreground screen component. I.e., a larger fraction of the dust is associated with the birthclouds, homogeneously mixed with the \Ha\ emitting regions. This finding too remains unchanged under the assumption of different scattering behaviors. It is reminiscent of the finding by \citet{daCunha2010} based on infrared observations that the fraction of dust luminosity contributed by the diffuse ISM drops with increasing specific SFR ($\sim \eqHa$ EW).

The best-fit $\beta$ in our model is positive, meaning that fraction of dust in a foreground screen increases for larger $R/R_e$ and is reduced in galaxy centres.  Based on an in depth (and higher resolution) analysis of M83 \citet{Liu2013} come to the same conclusion.  More generally, these authors detail how the physical structure of dust and emitting regions in M83 is more complex than any of the bracketing scenarios of a foreground screen or homogeneous mixture individually, and is best described by a combination of both, echoing our findings.  We note that the same picture of dual dust components underpins explanations for the observed differential extinction between gas and stars and its dependence on parameters such as inclination and specific SFR \citep{Charlot2000, Calzetti2000, Wild2011, Price2014, Reddy2015}.  For an application of the MaNGA data set to characterize this aspect of complex dust geometries at a spatially resolved level we defer the reader to \citet{Lininprep}.

The fact that our favored model (Figures\ \ref{5free.fig} to\ \ref{5free_residuals.fig}) still does not use stellar mass as an input observable yet reproduces at least qualitatively the enhanced extinction to spaxels in more massive galaxies reflects findings by \citet{Qinprep} based on galaxy-integrated SDSS + GALEX + WISE observations. These authors conclude that, once controlling for IR luminosity ($\sim$ SFR) and metallicity (setting the DGR) no explicit dependence on stellar mass is required to explain the observed IR/UV ratios. This suggests that any observed relation between extinction and stellar mass results indirectly from more massive galaxies featuring higher SFRs (the so-called star-forming main sequence) and higher metallicities (the well-known mass - metallicity relation).

\section{Summary and conclusions}
\label{summary.sec}

In this paper, we analyzed the resolved relation between star formation and extinction exploiting the exquisite number statistics of the MaNGA integral-field spectroscopic survey. Based on observations of 977 star-forming galaxies, together contributing measurements over 586459 spaxels, we establish empirically that the effective extinction $A_{\eqHa}$ and observed (dust-corrected) star formation surface density $\Sigma_{SFR}$ are tightly related, as $A_{\eqHa} = 0.46 \log(\Sigma_{SFR}) + 1.53$.  
Offsets from this relation show systematic dependencies on the gas-phase metallicity and \Ha\ EW of the spaxels, and inclination, size and mass of the galaxies they belong to. Spaxels with enhanced extinction tend to feature higher metallicities, lower EWs, and belong to larger, more massive and more inclined galaxies. Galactocentric distance and $\Sigma_{SFR}$ are strongly correlated but at a given $\Sigma_{SFR}$ the bulk of the spaxels show little variation in $R/R_e$ with extinction.

We present a simple model including an inverse star formation law to derive gas surface densities and adopting a dust-to-gas ratio which scales linearly with metallicity to compute the total dust column.  For a given parameterization of the dust geometry the latter can be translated to an effective extinction $A_{\eqHa}$. Gradually introducing extra complexity to our model, we find that the data require a dust geometry which is intermediate between a pure homogeneous mixture and a uniform foreground screen. Results consistent with the observations are obtained when assigning a minor ($\sim 12\%$) fraction of the dust to a clumpy foreground screen and the rest to a homogeneous mixture. Our best-fitting model further lets this breakdown between screen and mixture be a function of \Ha\ EW and galactocentric distance.

Irrespective of the latter nuances in implementation, the modeling favors a super-linear slope for the KS relation $n_{KS} \sim 1.47$. The model produces secondary dependencies consistent with the observations and residuals that are flat as a function of all observables considered. We conclude that the observed near-linear relation between $A_{\eqHa}$ and $\log(\Sigma_{SFR})$ can be understood at a spatially resolved level by connecting gas and dust scaling relations with a non-trivial dust geometry.

\acknowledgments
SW acknowledges support from the Chinese Academy of Sciences President's International Fellowship Initiative (grant no. 2017VMB0011). This project is supported by the National Natural Science Foundation of China under the grants No. 11603058. The authors wish to thank Caroline Bertemes for fruitful discussions.

Funding for the Sloan Digital Sky Survey IV has been
provided by the Alfred P. Sloan Foundation, the U.S.
Department of Energy Office of Science, and the Participating
Institutions. SDSS-IV acknowledges support and resources from
the Center for High-Performance Computing at the University of
Utah. The SDSS web site is \url{http://www.sdss.org}. SDSS-IV is
managed by the Astrophysical Research Consortium for the
Participating Institutions of the SDSS Collaboration including
the Brazilian Participation Group, the Carnegie Institution for
Science, Carnegie Mellon University, the Chilean Participation
Group, the French Participation Group, Harvard-Smithsonian
Center for Astrophysics, Instituto de Astrof\'isica de Canarias,
The Johns Hopkins University, Kavli Institute for the Physics
and Mathematics of the Universe (IPMU)/University of Tokyo,
Lawrence Berkeley National Laboratory, Leibniz Institut f\"ur
Astrophysik Potsdam (AIP), Max-Planck-Institut f\"ur Astronomie
(MPIA Heidelberg), Max-Planck-Institut f\"ur Astrophysik (MPA
Garching), Max-Planck-Institut f\"ur Extraterrestrische Physik
(MPE), National Astronomical Observatory of China, New
Mexico State University, New York University, University of
Notre Dame, Observat\'ario Nacional/MCTI, The Ohio State
University, Pennsylvania State University, Shanghai Astronomical Observatory, United Kingdom Participation Group, Universidad Nacional Aut\'onoma de M\'exico, University of Arizona,
University of Colorado Boulder, University of Oxford,
University of Portsmouth, University of Utah, University of
Virginia, University of Washington, University of Wisconsin,
Vanderbilt University, and Yale University.
%$ $\\

%\section*{appendix}
%\appendix
%\section{Appendix information}

%%%%%%%%%%%%%%%% %%%%%%%%%%%%%%%%%%
%\begin{figure}[t]
%\epsscale{0.8}
%\plottwo{eps/plot_onemodel_without_alphabeta1_residual_Z.eps}{eps/plot_onemodel_without_alphabeta1_residual_HaEW.eps}
%\plottwo{eps/plot_onemodel_without_alphabeta1_residual_RRe.eps}{eps/plot_onemodel_without_alphabeta1_residual_Ellipticity.eps}
%%\plottwo{eps/plot_onemodel_without_alphabeta1_residual_Re.eps}{eps/plot_onemodel_without_alphabeta1_residual_Mstar.eps}

%\newpage
%\clearpage

\bibliography{ms}

\end {document}